\documentclass[useAMS,usenatbib,fleqn]{mn2e}

\usepackage{caption}
\usepackage{graphicx}
\usepackage{amsmath}
\usepackage{amssymb}  
\usepackage{url}
\usepackage{times}

\def \aap{A\&A}

\def \apj{ApJ}
\def \apjl{ApJ}
\def \apjs{ApJS}
\def \araa{ARA\&A}

\def \jcap{J. Cos. Ast. Phys.}

\def \mnras{MNRAS}

\def \nat{Nat}

\def \pasa{Publ. Astron. Soc. Austral.}
\def \pasp{PASP}

\newcommand{\NHI}{\ensuremath{{\it N}_\mathrm{HI}}}

\newcommand{\lya}{\ensuremath{\rm Ly\alpha}}
\newcommand{\lyb}{\mbox{Ly$\beta$}}
\newcommand{\lyg}{\mbox{Ly$\gamma$}}
\newcommand{\kms}{\ensuremath{\rm km\,s^{-1}}}
\newcommand{\kmsMpc}{\ensuremath{\rm km\,s^{-1}\,Mpc^{-1}}}
\newcommand{\cmm}{\rm cm\ensuremath{^{-2}}}

\newcommand{\msun}{\ensuremath{{\it M}_\odot}}

\newcommand{\OmHI}{\ensuremath{\rm \Omega_{\rm HI}}}
\newcommand{\OmDLA}{\ensuremath{\rm \Omega^{\rm DLA}_{\rm HI}}}
\newcommand{\OmDLAg}{\ensuremath{\rm \Omega^{\rm DLA}_g}}

\newcommand{\AlII}{\hbox{{\rm Al}{\sc \,ii}}}

\newcommand{\CII}{\hbox{{\rm C}{\sc \,ii}}}

\newcommand{\HI}{\hbox{{\rm H}{\sc \,i}}}

\newcommand{\OI}{O{\sc\,i}}

\newcommand{\SiII}{\hbox{{\rm Si}{\sc \,ii}}}

\title[Neutral Hydrogen mass density at $z=5$]{The Neutral Hydrogen
  Cosmological Mass Density at $\mathbf{z=5}$}

\author[N. Crighton et al.]{Neil
  H. M. Crighton,$^1$\thanks{neilcrighton@gmail.com}
Michael T. Murphy,$^1$
J. Xavier Prochaska,$^2$
G\'{a}bor Worseck,$^3$ \newauthor
Marc Rafelski,$^4$
George D. Becker,$^5$
Sara L. Ellison,$^6$
Michele Fumagalli,$^{7,8}$
 \newauthor
Sebastian Lopez,$^9$ 
Avery Meiksin$^{10}$
and John M. O'Meara$^{11}$\\
\\
$^1$Centre for Astrophysics and Supercomputing, Swinburne University of Technology, Hawthorn, Victoria 3122, Australia\\
$^2$Department of Astronomy and Astrophysics, UCO/Lick Observatory, University of California, 1156 High Street, Santa Cruz, CA 95064, USA\\
$^3$Max-Planck-Institut f\"{u}r Astronomie, K\"{o}nigstuhl 17, D-69117 Heidelberg, Germany\\
$^4$Infrared Processing and Analysis Center, Caltech, Pasadena, CA 91125, USA \\
$^5$Space Telescope Science Institute, 3700 San Martin Dr, Baltimore, MD 21218, USA \\
$^6$Department of Physics and Astronomy, University of Victoria, Victoria, BC V8P 1A1, Canada\\
$^7$Institute for Computational Cosmology, Department of Physics, Durham University, South Road, Durham DH1 3LE, UK\\
$^8$Carnegie Observatories, 813 Santa Barbara Street, Pasadena, CA 91101, USA\\
$^9$Departamento de Astronom\'ia, Universidad de Chile, Casilla 36-D, Santiago, Chile\\
$^{10}$Scottish Universities Physics Alliance, Institute for Astronomy, University of Edinburgh, Blackford Hill, Edinburgh EH9 3HJ, UK\\
$^{11}$Department of Chemistry and Physics, Saint Michael's College, One Winooski Park, Colchester, VT 05439, USA\\
}

\begin{document}

\date{Accepted xxxx. Received xxxx; in original form xxxx}

\pagerange{\pageref{firstpage}--\pageref{lastpage}} \pubyear{xxxx}

\maketitle

\label{firstpage}

\begin{abstract}

We present the largest homogeneous survey of $z>4.4$ damped \lya\
systems (DLAs) using the spectra of 163 QSOs that comprise the Giant
Gemini GMOS (GGG) survey. With this survey we make the most precise
high-redshift measurement of the cosmological mass density of neutral
hydrogen, \OmHI. At such high redshift important systematic
uncertainties in the identification of DLAs are produced by strong
intergalactic medium absorption and QSO continuum placement. These can
cause spurious DLA detections, result in real DLAs being missed, or
bias the inferred DLA column density distribution. We correct for
these effects using a combination of mock and higher-resolution
spectra, and show that for the GGG DLA sample the uncertainties
introduced are smaller than the statistical errors on \OmHI. We find
$\OmHI=0.98^{+0.20}_{-0.18}\times10^{-3}$ at $\langle z\rangle=4.9$,
assuming a 20\% contribution from lower column density systems below
the DLA threshold. By comparing to literature measurements at lower
redshifts, we show that \OmHI\ can be described by the functional form
$\Omega_{\rm HI}(z)\propto(1+z)^{0.4}$. This gradual decrease from
$z=5$ to $0$ is consistent with the bulk of \HI\ gas being a
transitory phase fuelling star formation, which is continually
replenished by more highly-ionized gas from the intergalactic medium,
and from recycled galactic winds.
\end{abstract}


\section{Introduction}

The neutral hydrogen mass density of the universe, \OmHI, is an
important cosmological observable. It determines the precision with
which cosmological parameters can be constrained by observations of
the \HI\ intensity power spectrum
\citep[e.g.][]{Barkana07_IGM,Chang08_HI,Wyithe08_HI,Padmanabhan15},
and we expect its evolution to be linked to the cosmic star formation
history. The main contributor to \OmHI\ is high column density,
predominantly neutral gas clouds \citep[e.g.][]{OMeara07,Zafar13},
self-shielded from ionizing radiation and therefore likely fuel for
future star formation \citep[e.g.][]{Wolfe05}. Thus tracing the
evolution of \OmHI\ from the end of reionization, through the epoch of
the cosmic star formation peak at $z\sim2$ to the present day is of
central importance to our understanding of galaxy formation. It also
provides an excellent integral constraint against which theoretical
models of galaxy formation can be tested.

At redshift $<0.3$, \HI\ 21 cm emission can be used to measure
\OmHI\ either directly or by stacking analyses
\citep[e.g.][]{Zwaan05,Martin10_OmHI}. At higher redshifts, where
emission is too weak to be detected with current facilities,
\OmHI\ can instead be inferred from the incidence rate of damped
\lya\ systems (DLAs, defined as absorption systems with
$\NHI\ge20.3$~\cmm), which trace the bulk of neutral gas in the
universe \citep{Prochaska05_DLA}. These systems are detected in
absorption in the spectra of background QSOs, and their characteristic
damping wings allow column densities to be measured even at low
spectral resolution.

Early DLA surveys at $2<z<4$, which were typically comprised of a few
hundred QSOs and assumed a cosmological deceleration parameter
$q_0=0.5$ or $0$, suggested that the gas mass density in DLAs may have
been sufficient to produce most of the stars seen in the local
universe \citep{Lanzetta91dla,Wolfe95,StorrieLombardi96}. However, a
change to a modern concordance cosmology revealed that DLAs at
$z\sim3$ contain $<50$ percent of the present day mass density in
stars \citep[e.g.][see also Section
  \ref{s_prev}]{StorrieLombardi00,Peroux05}. In addition, recent DLA
surveys at $2<z<4$ using more than 10,000 QSOs assembled from the
Sloan Digital Sky Survey (SDSS)
\citep{Prochaska04_OmHI,Prochaska05_DLA,Prochaska09_OmHI,Noterdaeme09_OmHI,Noterdaeme12_OmHI}
have shown that there is very little evolution in the \HI\ mass
density from $z=3$ to the present day. This is starkly at odds with
the strong evolution in the star formation rate over the same period
\citep[e.g.][]{Madau14}.  One view is that \HI\ represents a
transitory phase fuelling star formation
\citep[e.g.][]{Prochaska05_DLA,Dave13}, which is continually
replenished by more highly ionized gas from either the intergalactic
medium (IGM) or recycled galactic outflows.

While it is important to constrain \OmHI\ across the whole of cosmic
history, it is of particular interest at the highest
redshifts. \citet{Rafelski14} report a decrease in the metal mass
density in damped \lya\ systems from $z=5$ to $4.5$, hinting at an
abrupt change in the enrichment of \HI\ gas past $z=5$. This may be
caused by a change in the population of objects containing neutral
hydrogen, which could be accompanied by a similarly abrupt evolution
in \OmHI. Moreover, since massive stars in galaxies are believed to
have reionized the Universe \citep[e.g.][]{Bouwens12}, it is important
to track the evolution of the fuel for star formation up to the epoch
of reionization. However, it is a challenge to assemble the large
sample of high-redshift QSO spectra necessary for a $z>4.5$ DLA
survey. The decline in the QSO space density at $z>3$ means that
relatively few redshift $>4.4$ QSOs were observed by the SDSS, and
those that were typically have too low a S/N to reliably identify
DLAs. For example, \citet{Rafelski12,Rafelski14} find a
misidentification rate of 26\% for DLA candidates from SDSS DR5 at
$z>4$, and of 97\% for candidates from DR9 at $z>4.7$. For this reason
smaller DLA surveys have been performed at higher redshift, often
using higher resolution spectra to make robust identifications of
DLAs. \citet{Peroux03}, \citet{Guimaraes09} and \citet{Songaila10}
have all presented measurements of \OmHI\ at
$z>4.5$. \citet[][hereafter S10]{Songaila10} give a cumulative result
including data from all these previous studies, and this represents
the highest redshift measurement of \OmHI\ to date. They use a sample
of 19 QSOs with emission redshifts $>4.5$, and their measurement hints
at a possible downturn in \OmHI\ at $z\ge4$, but the uncertainties
from sample variance at $z>4.3$ are large.

Here we measure \OmHI\ as traced by DLAs at $3.5<z<5.4$ using a
homogeneous sample of 163 QSOs with emission redshifts between 4.4 and
5.4. This represents an increase in redshift path of a factor of eight
over S10 at $z>4.5$. Identifying DLAs becomes increasingly difficult
at higher redshift, as \HI\ absorption from the highly-ionized
intergalactic medium (IGM) becomes more severe, and blending with
strong systems below the DLA threshold can cause misidentification of
DLAs. Therefore we carefully check for systematic misidentifications
in our sample using both mock spectra and higher resolution spectra of
DLA candidates. More than 70\% of our DLA candidates (and $>85\%$ at
$z>4.5$) have been observed at higher resolution
\citep{Rafelski12,Rafelski14}, allowing us to confirm their
\NHI\ despite the increased IGM blending at high redshift.

This paper is structured as follows. In Section~\ref{s_data} we
describe the QSO spectra used for the
analysis. Section~\ref{s_formalism} describes the formalism used to
derive \OmHI\ from our observations and Section~\ref{s_method}
describes our method for measuring the DLA incidence rate, accounting
for systematic effects. Section~\ref{s_OmHI} describes our main
result, a measurement of the neutral hydrogen mass density at $z=5$,
and discusses its implications. Section~\ref{s_summary} summarises our
conclusions. We assume a flat $\Lambda$CDM cosmology, with
H$_0=70\,\kms$Mpc$^{-1}$, $\Omega_{m,0}=0.3$ and
$\Omega_{\Lambda,0}=0.7$. All distances are comoving unless stated
otherwise. The data and code used for this paper are available at
\protect\url{https://github.com/nhmc/GGG_DLA}.

\section{Data}
\label{s_data}

\begin{figure*}
\includegraphics[width=1.40\columnwidth]{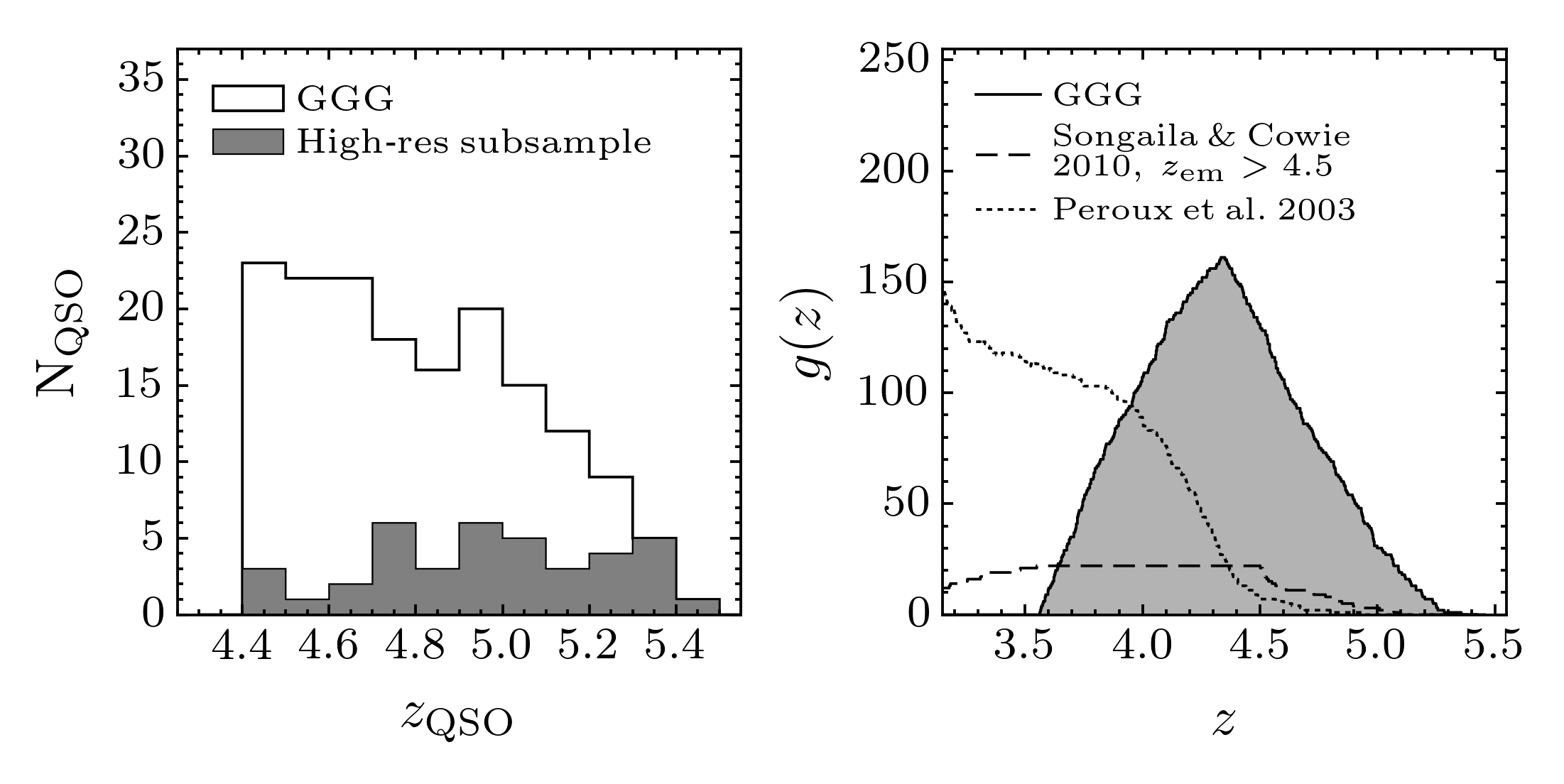}
\caption{\label{f_zpath} The left panel shows the emission redshift
  distribution for QSOs in the low-resolution GGG sample (open
  histogram), and for the subsample of these QSOs targeted with higher
  resolution spectra.  The right panel shows the redshift path,
  $g(z)$, for detecting DLAs for the GGG sample.  $g(z)$ is defined as
  the number of QSOs where a DLA can be detected as a function of DLA
  redshift. For comparison, $g(z)$ for previous high-redshift DLA
  surveys are also shown; for \citet{Peroux03} and for $z>4.5$ QSOs
  from S10. We do not show $g(z)$ for the SDSS DLA surveys
  \citep[e.g.][]{Noterdaeme12_OmHI}. Their $g(z)$ formally extends to
  $z>4$, but \citet{Prochaska05_DLA} warn this high redshift
  sensitivity should be viewed conservatively and
  \citet{Noterdaeme12_OmHI} do not include DLAs with $z>3.5$ in their
  statistical sample.}

\end{figure*}

Our main data sample consists of GMOS spectra for the 163 QSOs which
comprise the Giant Gemini GMOS (GGG) survey \citep{Worseck14_GGG}.
The QSOs were taken from the SDSS and all have emission redshifts
$4.4<z<5.4$. At these emission redshifts, the QSO sightlines are
likely unbiased regarding the number density of DLAs, unlike
sightlines with $2.7<z_\mathrm{em}<3.6$
\citep{Prochaska09_MFP,Worseck11_galex,Fumagalli13}. We also use a
smaller sample of 59 QSOs with higher resolution spectra, listed in
Table \ref{t_highres}. In contrast to the GGG sample, most of these
QSOs were targeted because of a known DLA candidate towards the
QSO. One of these higher resolution spectra was taken with the
Magellan Echellette Spectrograph on the Magellan Clay Telescope
\citep{Jorgenson13} and the remainder were taken with Echellette
Spectrograph and Imager on the Keck II Telescope \citep{Rafelski12,
  Rafelski14}. 39 of these QSOs are also in the GGG sample, and the
remaining 20 have a similar emission redshift to the GGG QSOs. We use
these higher resolution spectra to assess the reliability of our DLA
identifications and to estimate the importance of systematic effects,
but they are not included in the statistical sample used to measure
\OmHI. Figure \ref{f_zpath} shows the QSO emission redshift
distribution for our sample and the redshift path, $g(z)$, where DLAs
can be detected in comparison to previous high-redshift surveys. We
define
\begin{equation}
g(z) = \sum H(z^{\rm max}_i-z)H(z-z^{\rm min}_i)
\end{equation}
where $H$ is the heaviside step function, and $z^{\rm min}_i$ and
$z^{\rm max}_i$ are redshift limits for detecting DLAs in each QSO
spectrum \citep[e.g.][]{Zafar13}.

For a detailed description of the GGG spectra and the procedure used
to reduce them, see \citet{Worseck14_GGG}.  In brief, they were
observed with the Gemini Multi Object Spectrometers on the Gemini
telescopes, yielding a typical S/N $\sim20$ per 1.85~\AA\ pixel in the
\lya\ forest at a resolution of $\sim5.5$~\AA\ (full width at half
maximum, FWHM). The spectral coverage was tuned to be roughly constant
in the quasar rest frame (typically 850--1450~\AA). The
high-resolution ESI spectra we use\footnote{The reduced spectra are available
  at \protect{\url{http://www.rafelski.com/data/DLA/hizesi}}}
have a typical S/N of 15 per 10~\kms\ pixel and a resolution FWHM of
$31$~\kms\ (see Table~\ref{t_highres}). The single MagE spectrum has a
similar S/N but a resolution of $56$~\kms.

\section{Formalism}
\label{s_formalism}

Our aim is to measure the cosmic \HI\ mass density at $3.5<z<5.4$. The
bulk of the neutral gas at $2<z<5$ is in DLAs, with a $\sim 15\%$
contribution from sub-damped \lya\ systems (which have
$10^{19}<\NHI/(\cmm)<10^{20.3}$) and more highly ionized Lyman limit
and \lya\ forest absorbers with $\NHI<10^{19}\cmm$
\citep[][]{Peroux05,Prochaska05_DLA,OMeara07,Zafar13}. There are
several ways to express the comoving mass density of neutral hydrogen
used in the literature. For measurements at low redshift using radio
emission, authors typically quote $\Omega_\mathrm{HI}$, which is the
mass of neutral hydrogen alone, excluding any mass in molecules and
helium. For DLA absorption studies, authors generally quote the gas
mass in DLAs, $\Omega^\mathrm{DLA}_{g}$ (sometimes the $g$ subscript
is omitted) including a factor $\mu$ to account for
helium. \citet{Prochaska05_DLA} advocate using the quantity
$\Omega^\mathrm{neut}_g$, which is the mass in predominantly neutral
gas, which can be different from $\Omega^\mathrm{DLA}_{g}$. In this
work we quote the mass density from \HI\ alone, \OmHI, and exclude any
mass contribution from helium or molecules. Due to contamination and
the low resolution of the GMOS spectra, we only measure \HI\ in DLAs,
\OmDLA. To convert to \OmHI\ we apply a
correction derived from measurements of lower \NHI\ systems in
previous work.

We measure \OmDLA\ by counting the incidence rate of DLAs in the
spectra, and measuring \NHI\ from their strong damping wings. Below is
a summary of the formalism used to derive \OmDLA\ from the DLA
incidence rate. See section 4.1 of \citet{Prochaska05_DLA} and the
review by \citet{Wolfe05} for a more detailed description.

The number of DLAs in the intervals $(\NHI,\NHI+d\NHI)$ and $(X,X+dX)$
is defined as the frequency distribution, $f_\mathrm{DLA}(\NHI,X)d\NHI
dX$. Here $X$ is the `absorption distance', defined such that a
non-evolving population has a constant absorption frequency:
\begin{equation}
\label{e_X}
dX \equiv \frac{H_0}{H(z)} (1+z)^2\ dz
\end{equation}
where $H$ is the Hubble parameter. The DLA incidence rate is then
\begin{equation}
\ell_\mathrm{DLA}(X) dX = \int^\infty_{N_\mathrm{HI,min}} f_\mathrm{DLA}(\NHI,X)
d\NHI dX.
\end{equation}
It is related to the comoving number density of DLAs,
$n_\mathrm{DLA}(X)$, and the proper absorption cross section, $A(X)$,
by
\begin{equation} \label{e_n}
\ell_\mathrm{DLA}(X) = \frac{c}{H_0} n_\mathrm{DLA}(X) A(X).
\end{equation}

Since DLAs are mostly neutral, the \HI\ mass per DLA is
$m_\mathrm{H} \NHI A(X)$, where $m_\mathrm{H}$ is the hydrogen atom
mass. Combining this with equation~\ref{e_n} gives
\begin{equation}
\begin{split}
\Omega^\mathrm{DLA}_\mathrm{HI} (X) dX
&=\frac{H_0}{c}\frac{m_\mathrm{H}}{\rho_\mathrm{crit,0}}\int^\infty_{N_\mathrm{HI,min}}
\NHI f_\mathrm{DLA}(\NHI,X) d\NHI dX\\ &=\frac{8\pi
  G}{3H_0}\frac{m_\mathrm{H}}{c}\int^\infty_{N_\mathrm{HI,min}} \NHI
f_\mathrm{DLA}(\NHI,X) d\NHI dX.
\end{split}
\end{equation}
$N_\mathrm{HI,min}=10^{20.3}\cmm$, so this expression does not include
the contribution from lower \NHI\ systems to \OmHI. We discuss how we
include this contribution in section \ref{s_deltaHI}.

Due to the low resolution of the GMOS spectra, confusion from the
strong \lya\ forest absorption at $z>4$, uncertainty in the continuum
level, and systematics affecting sky subtraction, the measured
frequency of DLAs, $f_\mathrm{meas}(\NHI)$, may differ from the true
$f_\mathrm{DLA}$. Therefore we introduce a correction factor $k(\NHI)$
such that
\begin{equation}
f_\mathrm{DLA}(\NHI)=f_\mathrm{meas}(\NHI) k(\NHI).
\end{equation}
$k(\NHI)$ is the result of at least two effects. First, some systems
flagged as DLAs will actually be spurious (false positives), and some
real DLAs will be missed (false negatives). We estimate $k(\NHI)$ in
the following way. Let $N_\mathrm{cand}$ be the number of DLA
candidates flagged in our QSO survey. $N_\mathrm{cand,true}$ of these
candidates will be real DLAs, and the remainder will be spurious.  If
$N_\mathrm{true}$ is the true number of DLAs in the spectra, then we
can denote the fraction of DLA candidates which are not spurious as
$k_\mathrm{real} = N_\mathrm{cand,true} / N_{\rm cand}$, and the
fraction of true DLAs that are correctly identified as
$k_\mathrm{found} = N_\mathrm{cand,true} / N_\mathrm{true}$. This
gives
\begin{equation}
f_\mathrm{DLA}(\NHI)=f_\mathrm{meas}(\NHI)
\frac{k_\mathrm{real}}{k_\mathrm{found}}
\end{equation}
and thus $k(\NHI) = k_\mathrm{real}/k_\mathrm{found}$. In the
following sections we describe how we measure $f_\mathrm{meas}$, and
how high-resolution and mock spectra are used to estimate
$k_\mathrm{real}$ and $k_\mathrm{found}$.

\subsection{Other systematic effects contributing to $k(\NHI)$}

In measuring $k(\NHI)$ we explicitly take into account the rate of
spurious DLAs (false positives) and missed DLAs (false negatives).
There are several other systematic effects which could also contribute
to $k(\NHI)$, which we discuss here.

The first of these is any uncertainty in the \NHI\ measurements. If
there are large uncertainties in \NHI, or systematic offsets in the
\NHI\ estimated from the spectra as a function of \NHI, this may
change the inferred $f(\NHI)$.  However, in section \ref{s_N_z} we
show that the \NHI\ error from the GMOS spectra ($0.2$ dex) does not
have a detectable systematic bias, and section \ref{s_OmHI} shows that
any errors it introduces to \OmHI\ are negligible compared to other
uncertainties.  A related effect is for \NHI\ measurements at the DLA
threshold of \NHI$=10^{20.3}$~\cmm, where the more numerous lower
column density systems may be counted as DLAs through
\NHI\ uncertainties. This bias is a net source of false positives, and
so should be taken into account by our procedure for estimating
$k_\mathrm{real}$.

A second possibility is the presence of dust in DLAs. If DLAs contain
large amounts of dust they are able to extinguish the light from a
background QSO, removing these sightlines from our survey. In this
case we would measure a lower incidence of high metallicity, high
\NHI\ DLAs, which presumably contain the most dust. However, several
studies have shown that most DLAs are not associated with significant
amounts of dust (e.g.  \citealt{Murphy04}, \citealt{Vladilo08}), and
DLAs towards radio-selected QSOs, which are insensitive to the
presence of dust, have a similar \NHI\ distribution to those in
optically-selected QSOs \citep{Ellison01,Jorgenson06}.
\citet{Pontzen09} find that the cosmic \HI\ mass density may be
underestimated by $3$--$23\%$ at $z\sim3$ due to selection biases from
dust. We do not include this relatively small effect in our analysis,
but note where its inclusion would affect our conclusions.

Gravitational lensing may also introduce a bias. DLA host galaxies may
lens background QSOs, making them more likely to be found in our
survey. This would result in brighter QSOs being more likely to show
foreground DLA absorption compared to fainter QSOs. At $z\sim3$,
\citet{Murphy04} found evidence at the $\sim2\sigma$ level that DLAs
tend to be found towards brighter QSOs. \citet{Prochaska05_DLA} found
a higher incidence rate of high \NHI\ DLAs towards brighter QSOs
compared to fainter QSOs over a redshift range 2--4.5, that resulted
in a significant ($>95$ percent) difference in \OmHI\ between the two
samples. They attributed this effect to gravitational lensing. We
confirm that this effect is also present in our sample (which has some
overlap with the Prochaska et al. sample): there is a $25\pm15$
percent higher incidence rate of DLAs towards QSOs with $z$-band
magnitude $\le 19.2$ compared to QSOs with $z>19.2$~mag. DLAs towards
bright QSOs also tend to have high \NHI, resulting in a 30 percent
increase in \OmHI\ for the brighter compared to the fainter QSO
sample. The significance of the excesses we measure is modest
($1.7\sigma$), and a Kolmogorov-Smirnov test between the
\NHI\ distributions towards $z\le19.2$ and $z>19.2$~mag quasars yields
$D=0.3$ and a probability of 22\% that the two samples are drawn from
the same underlying distribution. Therefore, while this difference
hints at a selection effect related to the background QSO brightness,
we cannot yet rule out a simple statistical fluctuation.  We further
discuss how this possible bias may affect our \OmHI\ measurement in
Section \ref{s_is_lensing}.

\subsection{Conversion from \OmDLA\ to \OmHI}
\label{s_deltaHI}

Previous absorption studies have shown that the dominant contribution
to \OmHI\ is from DLAs. Lower column density systems also contribute
an appreciable fraction of \OmHI, however. This fraction is
$15$--$30$\% at $z=3$, depending on the assumed \NHI\ distribution
\citep[e.g.][]{OMeara07,Noterdaeme09_OmHI,Prochaska10,Zafar13}. To
parametrize this uncertainty, we introduce a correction factor
$\delta_{\rm HI}\equiv\OmHI/\OmDLA$ to convert between \OmDLA, which
we measure, and \OmHI.  We assume the \NHI\ distribution at $z>4$ is
not dramatically different from that at $z\sim3$ and take
$\delta=1.2$, which implies a 20\% contribution from lower column
density systems.  Zafar et al. find the contribution of sub-damped
systems to \OmHI\ increases with redshift, possibly due to a weakening
of the UV background as the number density of QSOs drops at high
redshift. Therefore a goal of future surveys should be to measure the
contribution of these sub-damped systems at $z > 4$.

\section{Method}
\label{s_method}

\subsection{Procedure for identifying DLAs}
\label{s_procedure}

We measure the frequency of DLAs, $f_\mathrm{meas}$, by identifying
DLA candidates by eye in the GMOS spectra, and then correcting for any
biases in identification using mock spectra. To identify candidates we
performed the following steps for each QSO spectrum:

\begin{enumerate}
\item Estimate the continuum as a spline, placing the spline knot
  points by hand. We used the low-$z$ composite QSO spectrum from
  \citet{Shull12} to indicate the position of likely QSO emission
  lines which fall inside the \lya\ forest.
\item Look for a possible damped \lya\ line in the \lya\ forest
  between the QSO \lya\ and \lyb\ emission lines. Estimate its
  redshift and \NHI\ by plotting a single component Voigt profile with
  $b=30$~\kms\ over the spectrum, and varying \NHI\ and $z$ until it
  matches the data by eye\footnote{This $b$ value was chosen for
    convenience. The precise $b$ used does not strongly affect the
    \lya\ profile.}. If necessary the continuum was varied at the same
  time \NHI\ was estimated to obtain a plausible fit. At higher
  redshifts, blending with IGM absorption can make estimating
  \NHI\ challenging, as the damping wings can be very heavily blended
  with IGM absorption. In this case the best constraint on \NHI\ is
  not from the shape of the damping wings, but instead from the extent
  of the \lya\ trough consistent with zero flux, and from any
  higher-order Lyman transitions.
\item If a candidate DLA is found based on the \lya\ profile, use its
  higher-order Lyman series (if available in the spectrum) to refine
  its redshift and \NHI.
\item Repeat steps (ii) \& (iii) for all DLA candidates in the
  \lya\ forest.
\end{enumerate}

DLA absorption can also be detected bluewards of the QSO
\lyb\ emission line. However, we chose to search only between
\lya\ and \lyb\ emission in our sample to maximise the chance of
having useful Lyman series lines in addition to \lya, and to avoid any
additional systematic effects caused by further blending with the
\lyb\ forest. While most DLAs also have associated metal lines
detected by the GMOS spectra, we did not use any metal line
information when measuring the DLA candidate redshift or \NHI. This
was done to avoid any bias against finding low metallicity systems,
which may not have detectable metals in the GMOS spectra.

\begin{figure*}
\begin{tabular}{c}
\includegraphics[width=0.7\textwidth]{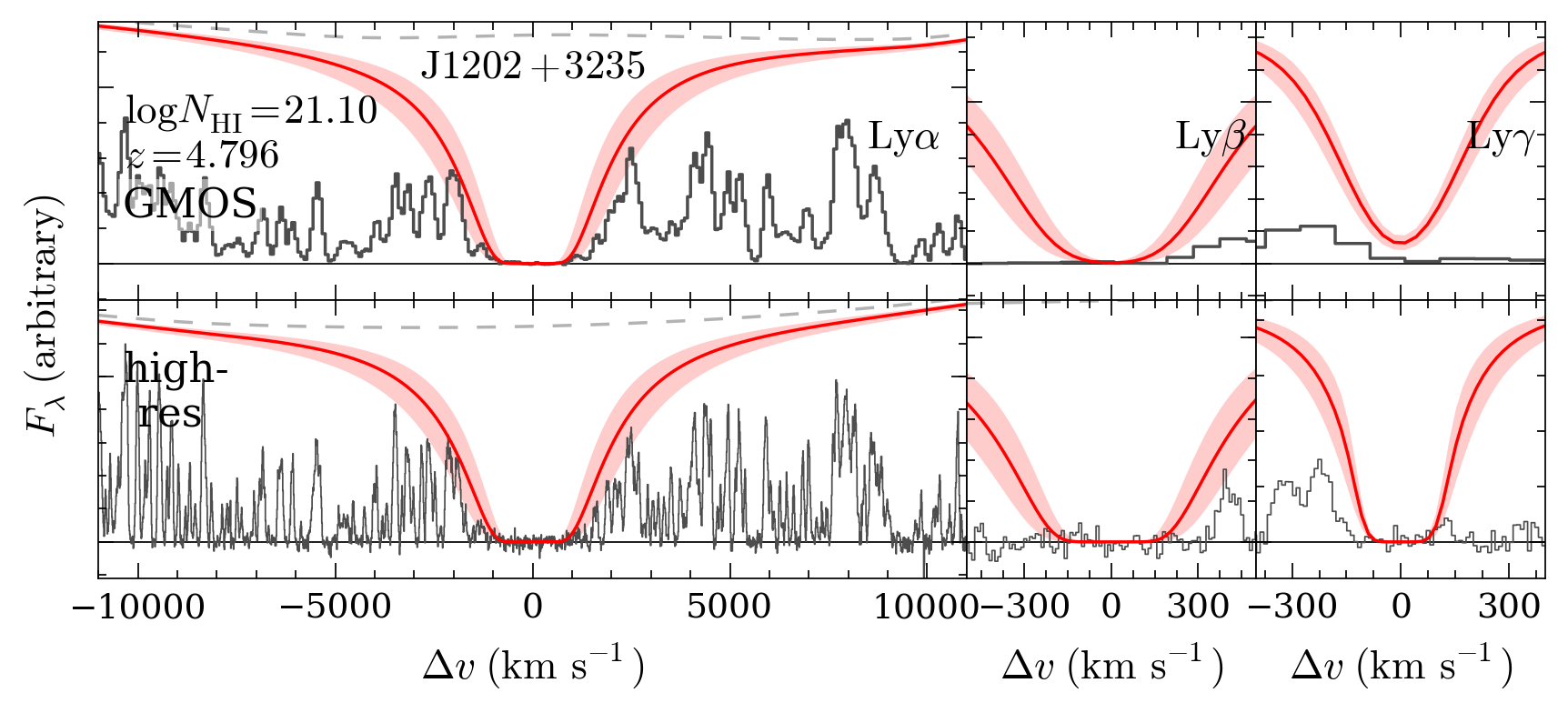} \\
\includegraphics[width=0.7\textwidth]{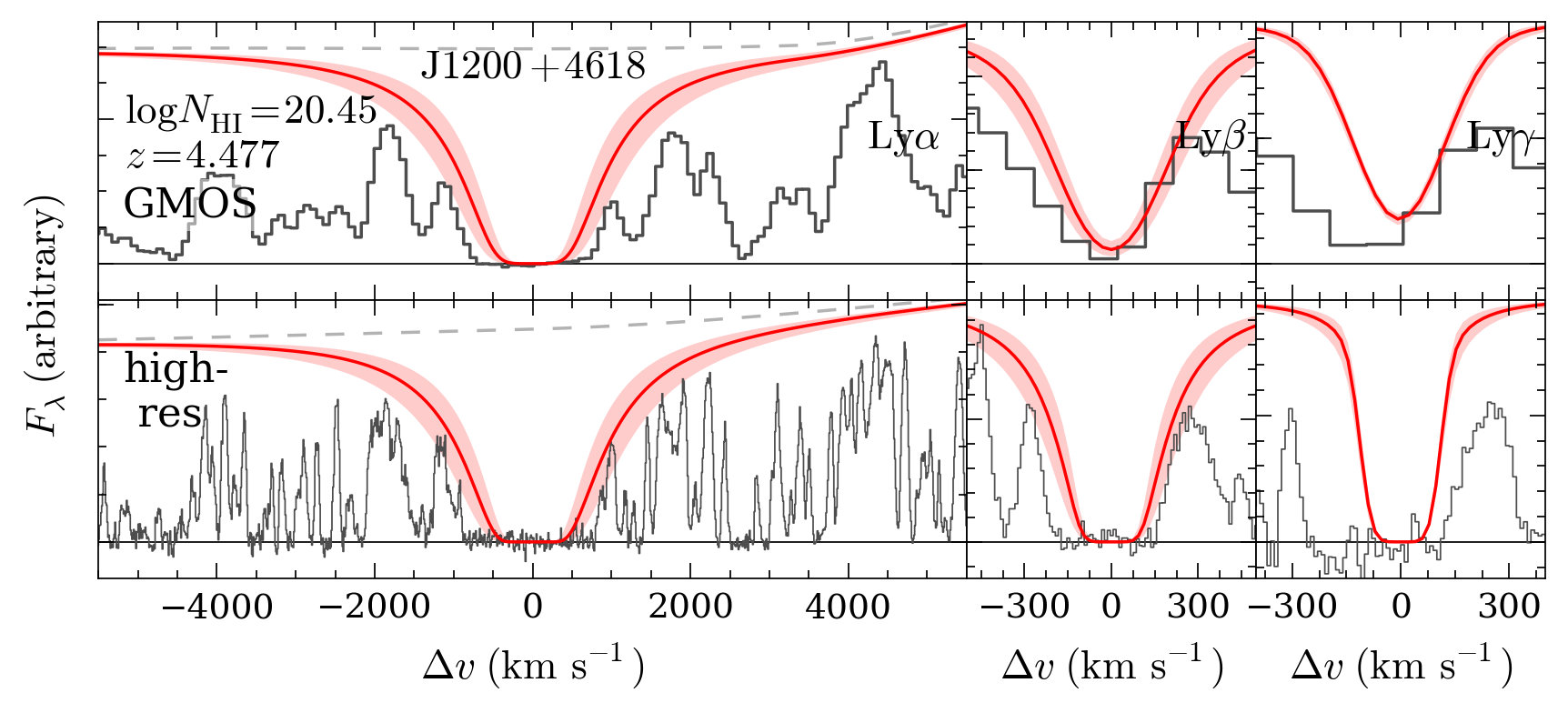}
\end{tabular}
\caption{\label{f_correct} DLAs identified in the GMOS spectra
  (resolution FWHM $\sim230$~\kms) which are confirmed in higher
  resolution ESI spectra (resolution FWHM $\sim30$~\kms). In each case
  the top panels show the GMOS spectrum and the bottom panels the ESI
  spectrum of the same QSO. The model shows the \NHI\ and redshift
  estimated from the ESI spectra with the redshift fixed by low-ion
  metal absorption. The shaded region shows an uncertainty in
  $\log\NHI$ of 0.2. The \NHI\ and redshift estimated from the GMOS
  spectra are given in Table~\ref{t_dla}.}
\end{figure*}

Two of the authors (NHMC and JXP) searched the spectra for DLAs
independently. The above steps were done either using custom-written
Python code, or with \textsc{x\_fitdla} from \textsc{xidl}, depending
on which author performed the search. For each QSO we also noted any
properties of the spectrum which might complicate the identification
of DLAs, such as the presence of broad absorption lines associated
with the background QSO, or of possible problems with the sky
background subtraction. Two example DLA candidates are shown in
Figure~\ref{f_correct}. In these two cases, higher resolution spectra
confirm that both candidates are indeed DLAs. The \NHI\ and redshift
estimated from the GMOS spectra differ slightly from the values
inferred from the higher resolution spectra -- we discuss this issue
further in Section~\ref{s_N_z}. Once we assembled a list of DLA
candidates, we selected only those within a redshift path limit
defined by:
\begin{equation}
\begin{split}
z_\mathrm{min} &= (1 + z_\mathrm{qso}) \frac{\lambda_\mathrm{Ly\beta}}{\lambda_\mathrm{Ly\alpha}} - 1 \\
z_\mathrm{max} &= (1 + z_\mathrm{qso}) (1 - \delta v / c) - 1
\end{split}
\end{equation}
where $\lambda_\mathrm{Ly\alpha}=1215.6701$~\AA,
$\lambda_\mathrm{Ly\beta}=1025.72$~\AA\ and $\delta v =
5000\,\kms$. This $\delta v$ was chosen to exclude `proximate' DLAs,
whose incidence rate is likely affected by a combination of ionizing
radiation from the background QSO, and by the overdensity associated
with the QSO host galaxy halo
\citep[e.g.][]{Ellison02,Russell06,Prochaska08_PDLA,Ellison10}.
Table~\ref{t_zpath} lists the redshift path limits used for each QSO
in the GGG sample. We then convert the redshift path for each QSO to
an absorption distance path using equation~(\ref{e_X}).

With these DLA candidate lists we can derive the measured incidence
rate of DLAs, $f_\mathrm{meas}$. However, despite our attempt to take
continuum uncertainties and IGM absorption into account when measuring
\NHI\ for each DLA, large systematic uncertainties may remain.  The
following sections describe how we quantify these uncertainties using
the correction factors $k_\mathrm{real}$ and $k_\mathrm{found}$ to
$f_\mathrm{meas}$.

\subsection{Estimation of $k_\mathrm{real}$ and $k_\mathrm{found}$}

We expect $k_\mathrm{real}$ to be less than unity, meaning that there
are some spurious DLA candidates. The rate of these spurious
candidates is estimated in two ways. First, we use the sample of
higher-resolution spectra to identify DLAs, and compare these with the
DLA candidates found in the low-resolution sample. Second, we create
mock low-resolution spectra which closely match the GMOS spectra and
contain DLAs generated from a distribution at $z=3$, and then search
these spectra for DLAs in the same way as the real spectra.

$k_\mathrm{found}$ is also expected to be less than unity, which means
some true DLAs exist which we do not flag as DLA candidates in the low
resolution spectra. Again we estimate the fraction of true DLAs
recovered in two independent ways, using higher resolution spectra and
mocks. In the first case DLAs identified in the higher resolution QSO
spectra were used as a reference list of true DLAs, and compared to
the candidate DLAs found in the lower resolution spectra of the same
QSOs. In the second case we used mock GMOS spectra, which allow us to
directly compare known DLAs in the spectra to the DLA candidates.

Our motivation for using two different ways to estimate the
correction factors (mocks and high resolution spectra) is to test
different systematic effects.  The main advantage of the mocks is that
the true DLA properties are known precisely. However, while we attempt
to reproduce the real spectra as closely as possible, including
\lya\ forest clustering, QSO redshift and signal-to-noise
distribution, it is still possible that the mocks may differ from the
real GMOS spectra. Metal absorption (not included in the mocks) or
clustering of strong absorbers that is different to the mocks may
cause more spurious DLAs. Alternatively, non-Gaussian noise in the
real spectra at low fluxes may mean that true DLAs are more likely to
be missed in the real spectra. Conversely, for the high-resolution
sample the true DLA properties are not known with complete certainty,
but the correct clustering, IGM blending, noise and metal absorption
are all included. Therefore these two approaches provide complementary
estimates of $k_\mathrm{found}$ and $k_\mathrm{real}$. The following
sections describe these approaches in more detail.

\subsubsection{Corrections using high resolution spectra}

DLAs can be found more easily in our sample of high resolution
spectra, and their \NHI\ and redshift are more accurately measured, in
comparison to the lower resolution GMOS spectra. Therefore we
independently identify DLAs in these spectra for the purpose of
deriving the correction factors $k_\mathrm{real}$ and
$k_\mathrm{found}$, and to test for any systematics in estimating
\NHI\ and $z$ for each DLA.  When identifying the DLAs in the 59
high-resolution spectra we follow the same process outlined for the
lower-resolution spectra in Section~\ref{s_procedure}, using the Lyman
series to estimate the redshift and \NHI. However, we also refine the
redshift and \NHI\ using the position of low-ionization metal lines
(\OI, \SiII, \CII\ and \AlII) where possible.  For the 20 QSOs with
high-resolution spectra which are not in the GGG sample, we created
low-resolution spectra by convolving the high-resolution spectra to
the same FWHM resolution, and rebinning to the same pixel size as the
GMOS spectra. The same noise array was used for these spectra as for
the GGG QSO with a redshift closest to each QSO, normalising such that
the median S/N within rest-frame wavelengths
$1260$--$1280$~\AA\ match. These low resolution spectra were searched
for DLAs in the same way as the GMOS spectra.

In this way we made two lists of DLAs, one from the high resolution
spectra, and another from low-resolution spectra of the same QSOs. The
DLAs identified in the higher resolution sample are listed in columns
5 and 6 of Table~\ref{t_dla}. We then estimated $k_\mathrm{real}$ as
$N_\mathrm{cand,true}/N_\mathrm{cand}$, where $N_\mathrm{cand}$ is the
number of DLA candidates from the low-resolution spectra, and
$N_\mathrm{cand,true}$ is the number of those candidates confirmed to
be DLAs by the high resolution spectra. $k_\mathrm{found}$ is
estimated as $N_\mathrm{cand,true}/N_\mathrm{true}$, where
$N_\mathrm{true}$ is the number of DLAs found in the high-resolution
spectra and $N_\mathrm{cand,true}$ is the number of those also flagged
as DLA candidates in the low resolution spectra. We calculate the
binomial confidence intervals on $k_\mathrm{real}$ and
$k_\mathrm{found}$ using the method described by \citet{Cameron11}.

\begin{figure*}
\includegraphics[width=0.7\textwidth]{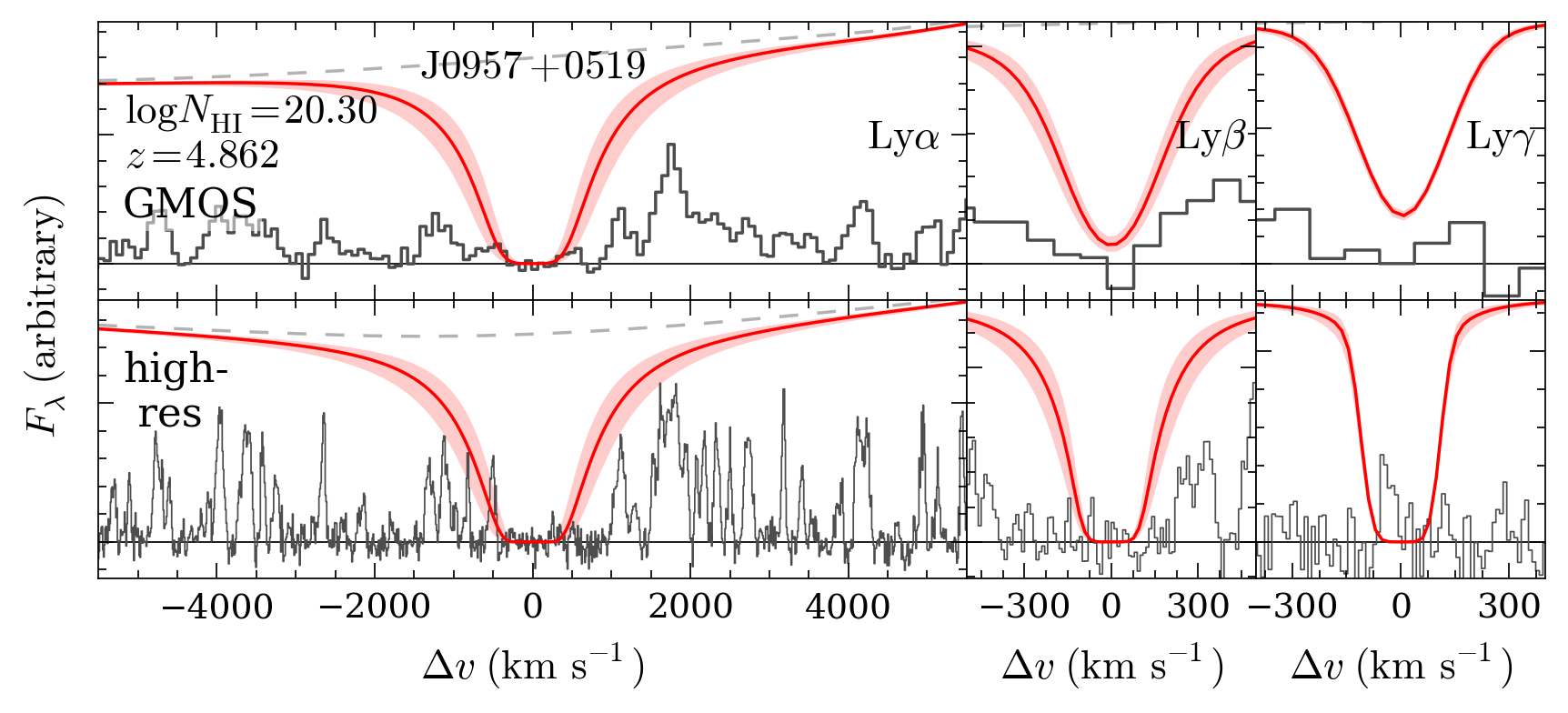}
\caption{\label{f_spurious} Example of a spurious DLA candidate. This
  was identified as a DLA with $\NHI=10^{20.4\pm0.2}$~\cmm\ in the
  GMOS spectrum shown in the top panels. However, the residual flux
  spikes at \lya\ and \lyg\ in the higher resolution (FWHM
  $\sim30$~\kms) ESI spectrum in the bottom panels show this system
  must have $\NHI<10^{20.3}$~\cmm.}
\end{figure*}

\begin{figure*}
\includegraphics[width=0.7\textwidth]{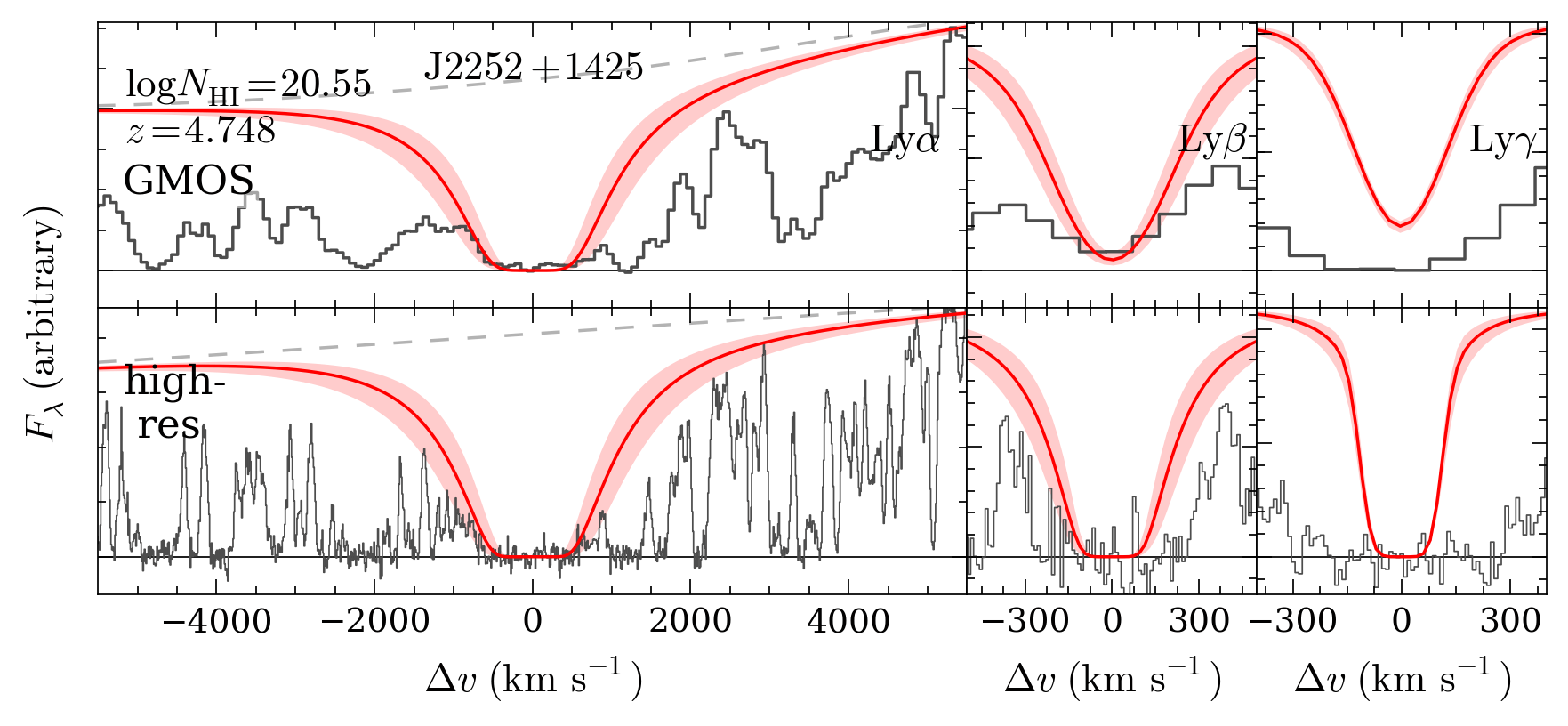}
\caption{\label{f_missed} A DLA that was not correctly identified in
  the GMOS spectra. Lower panels show the DLA in the ESI spectrum,
  with $\NHI=10^{20.55\pm0.2}$~\cmm. The residual flux in the core of
  the \lya\ line in the GMOS spectrum, however (top left panel), meant
  this system was missed. This residual flux around velocities $\Delta
  v \sim0$~\kms\ may be caused by either statistical fluctuations or
  systematics associated with sky background level.}
\end{figure*}

With this procedure we find $k_\mathrm{real}= 0.80_{-0.08}^{+0.07}$
and $k_\mathrm{found}=0.84^{+0.06}_{-0.08}$ using DLAs identified by
JXP (see Figures \ref{f_kfound}, \ref{f_kreal}) with similar values
found by NHMC. Both are below unity, and so there are both spurious
DLA candidates, and real DLAs missed.  Spurious DLAs usually occur
when flux spikes are smoothed away at GMOS resolution, making a lower
\NHI\ system appear to have strong damping wings.  An example spurious
DLA is shown in Figure \ref{f_spurious}.
Real DLAs are generally missed due to flux fluctuations in the core of
the \lya\ line: an example is shown in Figure \ref{f_missed}.

\subsubsection{Corrections using mock spectra}

Our method for generating mock spectra is described in
Appendix~\ref{a_mocks}. In this case the \NHI\ for each DLA is known,
and so can be directly compared to the candidates identified in the
low-resolution mocks. Again $k_\mathrm{real}$ is estimated as
$N_\mathrm{cand,true}/N_\mathrm{cand}$, where $N_\mathrm{cand}$ is the
number of DLA candidates from the low-resolution mock spectra, and
$N_\mathrm{cand,true}$ is the number of those candidates that are
DLAs. $k_\mathrm{found}$ is estimated as
$N_\mathrm{cand,true}/N_\mathrm{true}$, where $N_\mathrm{true}$ is the
true number of DLAs in the mocks and $N_\mathrm{cand,true}$ is the
number of those recovered as DLA candidates. Again we calculate the
errors on $k_\mathrm{real}$ and $k_\mathrm{found}$ assuming a binomial
confidence interval. For the mocks we find
$k_\mathrm{real}=0.71\pm0.06$ and
$k_\mathrm{found}=0.92^{+0.04}_{-0.07}$ using DLAs identified by JXP
(see Figures \ref{f_kfound}, \ref{f_kreal}). Similar values are found by NHMC
(see Figures \ref{f_kfound_hiN}, \ref{f_kreal_hiN}).

\subsubsection{Comparison of correction factors and their dependence on redshift and column density}
 
We expect $k_\mathrm{real}$ and $k_\mathrm{found}$ to be a function of
a DLA's \NHI\ (high \NHI\ candidates should be more reliable),
spectral S/N (low S/N spectra will produce more spurious candidates)
and redshift (more spurious DLAs will be found at high redshift where
there is more IGM absorption). The most important of these for our
measurement of \OmHI\ is any redshift or
\NHI\ dependence. \citet{Noterdaeme09_OmHI} and
\citet{Noterdaeme12_OmHI} show that at $z\sim2.5$, systems with
$\NHI=10^{20.6-21.5}~\cmm$ make the largest contribution to
\OmHI. Thus we expect completeness corrections in this column density
range to have the largest effect on the final derived
\OmHI.\footnote{Due to our relatively small DLA sample, we may be
  missing some very high \NHI\ systems with
  $\NHI\ge10^{22}$~\cmm. These contribute only $10\%$ of \OmHI\ at
  $z\sim3$ \citep{Noterdaeme12_OmHI} and thus we do not expect their
  absence from our sample to strongly bias our results.}

\begin{figure}
\begin{tabular}{c}
\includegraphics[width=0.95\columnwidth]{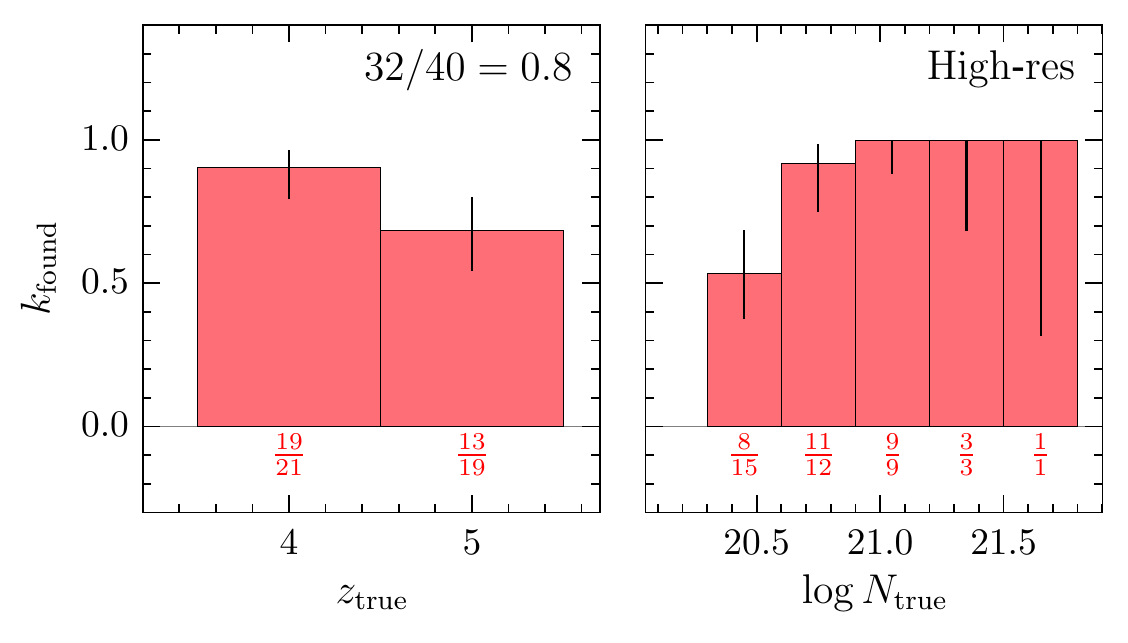} \\
\includegraphics[width=0.95\columnwidth]{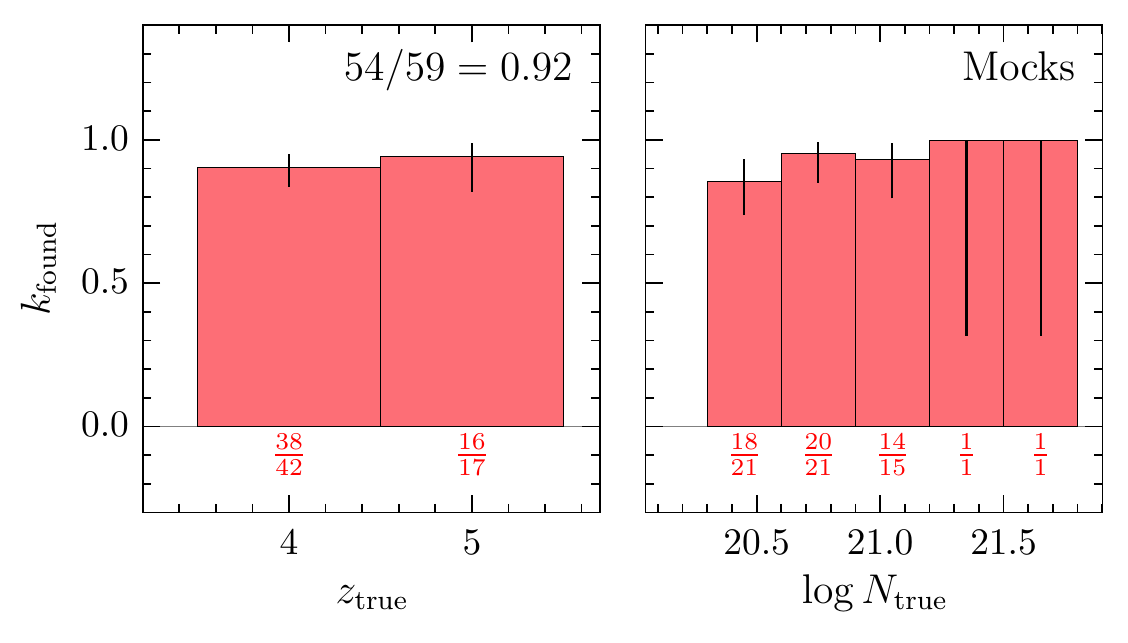}
\end{tabular}
\caption{\label{f_kfound} The fraction of true DLAs that were
  correctly identified by one of the authors (JXP),
  $k_\mathrm{found}$, as a function of the true redshift and \NHI. Top
  panels are for the high-resolution sample, bottom panels are for
  mocks. The upper row of numbers under each histogram gives the
  number of DLA candidates that are correct per bin, the lower row the
  total number of candidates. The total numbers for all bins are given
  at the top in the left panels. Vertical lines show the binomial
  $68\%$ uncertainties.}
\end{figure}

\begin{figure}
\begin{tabular}{c}
\includegraphics[width=0.95\columnwidth]{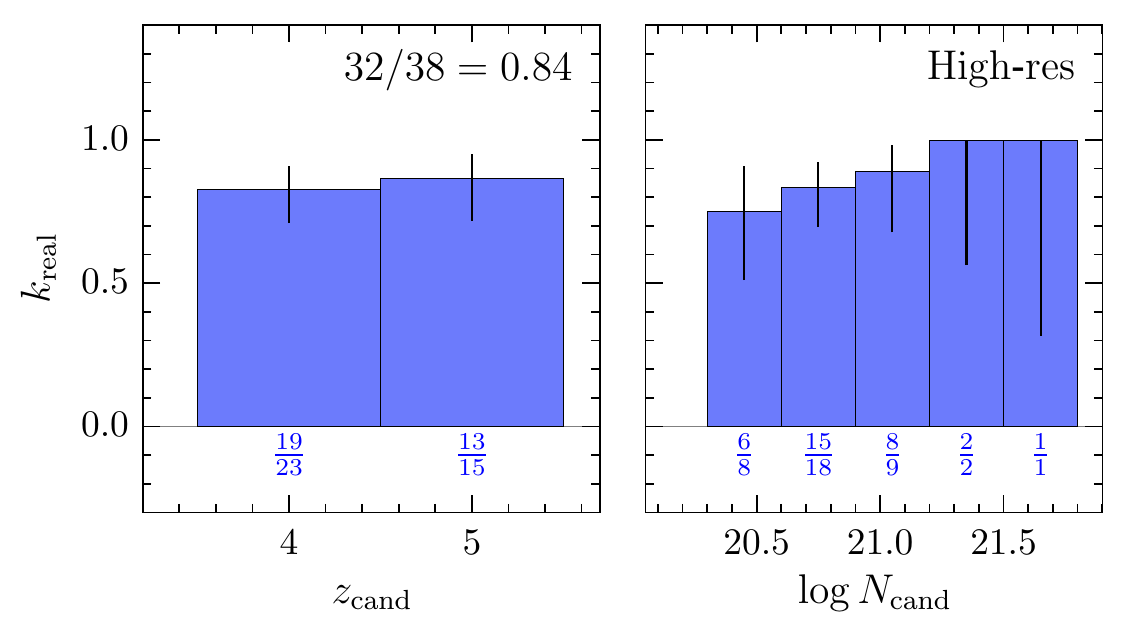} \\
\includegraphics[width=0.95\columnwidth]{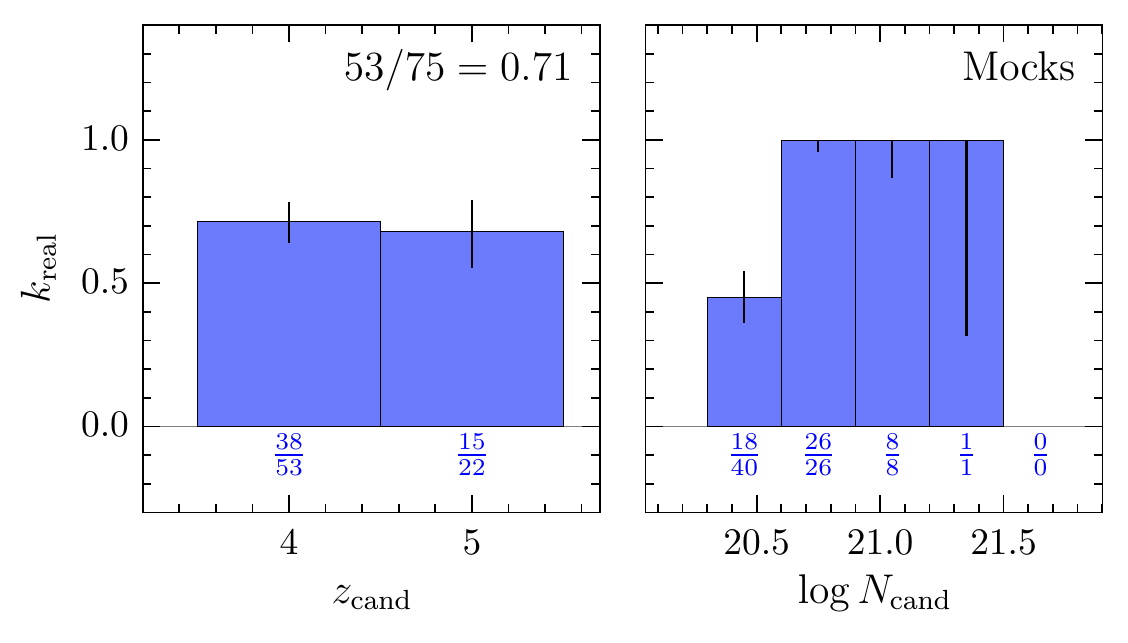} 
\end{tabular}
\caption{\label{f_kreal} The fraction of non-spurious DLA candidates,
  $k_\mathrm{real}$ by one of the authors (JXP), as a function of the
  candidate redshift and \NHI. Top panels are for the high-resolution
  sample, bottom panels are for mocks. The upper row of numbers under
  each histogram gives the number of true DLAs that are recovered per
  bin, the lower row the total number of true DLAs. The total numbers
  for all bins are given at the top of the left panels. Vertical lines
  show the binomial $68\%$ uncertainties.}
\end{figure}

The top panels of Figure~\ref{f_kfound} show the correction factor
$k_\mathrm{found}$ from the high-resolution spectra binned by the true
DLA redshift and \NHI, and the bottom panels show the same correction
factor estimated from the mocks. Figure \ref{f_kreal} shows the
correction factor $k_\mathrm{real}$ binned by the candidate DLA
redshift and \NHI, again for the high-resolution spectra and
mocks. These are derived from DLAs identified by one of the authors
(JXP) who search the spectra for DLAs, but values for the other author
(NHMC) are similar. There is no evidence for a strong dependence of
$k_\mathrm{real}$ or $k_\mathrm{found}$ on redshift, using either the
high-resolution spectra or the mocks. However, there is a weak
dependence of $k_\mathrm{real}$ and $k_\mathrm{found}$ on \NHI, with
the lowest \NHI\ bin having a significantly lower $k_\mathrm{real}$
than for higher \NHI\ bins. This matches our expectations: weaker
candidate DLAs are more likely to be spurious, and true DLAs that are
weak are more likely to be missed. We take this \NHI\ dependence into
account when applying the correction factors as described in
Section~\ref{s_OmHI}. We find no strong dependence of the correction
factors on S/N in either the mocks or the high-resolution sample for
the range of S/N the GMOS spectra cover.
 
Figures~\ref{f_kfound} and \ref{f_kreal} also show that corrections
derived from the mocks and high-resolution spectra are in reasonable
agreement. The main difference is in the number of spurious systems
with $\NHI\sim10^{20.3-20.6}$~\cmm. The right hand panels of
Figure~\ref{f_kreal} show that there are more weak, spurious DLAs
found in the mocks compared to the real GMOS spectra. However, we show
in the following section that the correction factor in this
\NHI\ range is not important for estimating \OmHI, and for the
remaining bins the mocks and high-resolution corrections match to
within $20$ per cent. As we discussed earlier, the high-resolution
sample and mocks test different systematic uncertainties which may
affect \OmHI. Therefore the consistency of the correction factors
between these two methods suggests the mocks reproduce the true GMOS
spectra well, and that DLAs have been identified correctly in the
higher-resolution spectra.

\subsection{Uncertainties in \NHI\ and redshift}
\label{s_N_z}

If DLA column densities estimated from the GMOS spectra are
systematically in error, our measurement of \OmHI\ may be biased. Such
a systematic could occur because of incorrect placement of the
continuum, or blending of damping wings with the \lya\ forest.  This
is an additional effect not accounted for by the correction factor,
$k$, to $f_\mathrm{meas}$. Therefore, we search for any systematic
offset in \NHI\ by matching DLA candidates from the low-resolution
spectra to known DLAs in the high-resolution sample and mocks.

\begin{figure}
\begin{tabular}{c}
\includegraphics[width=1.01\columnwidth]{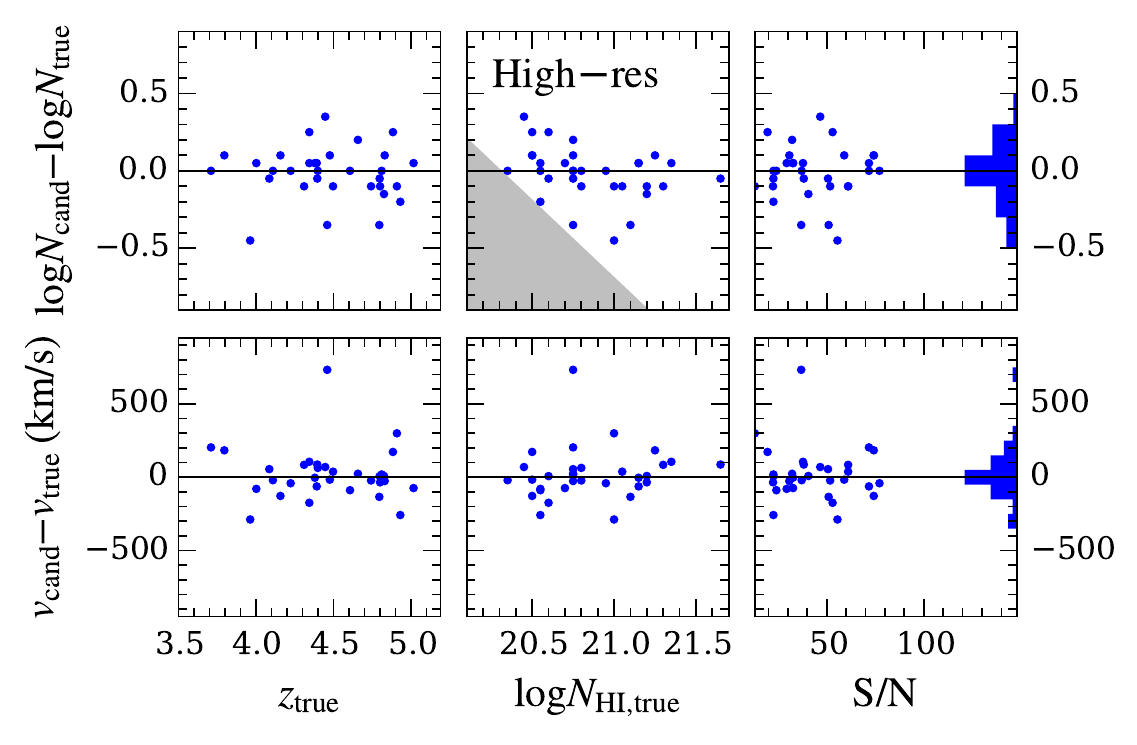} \\ 
\includegraphics[width=1.01\columnwidth]{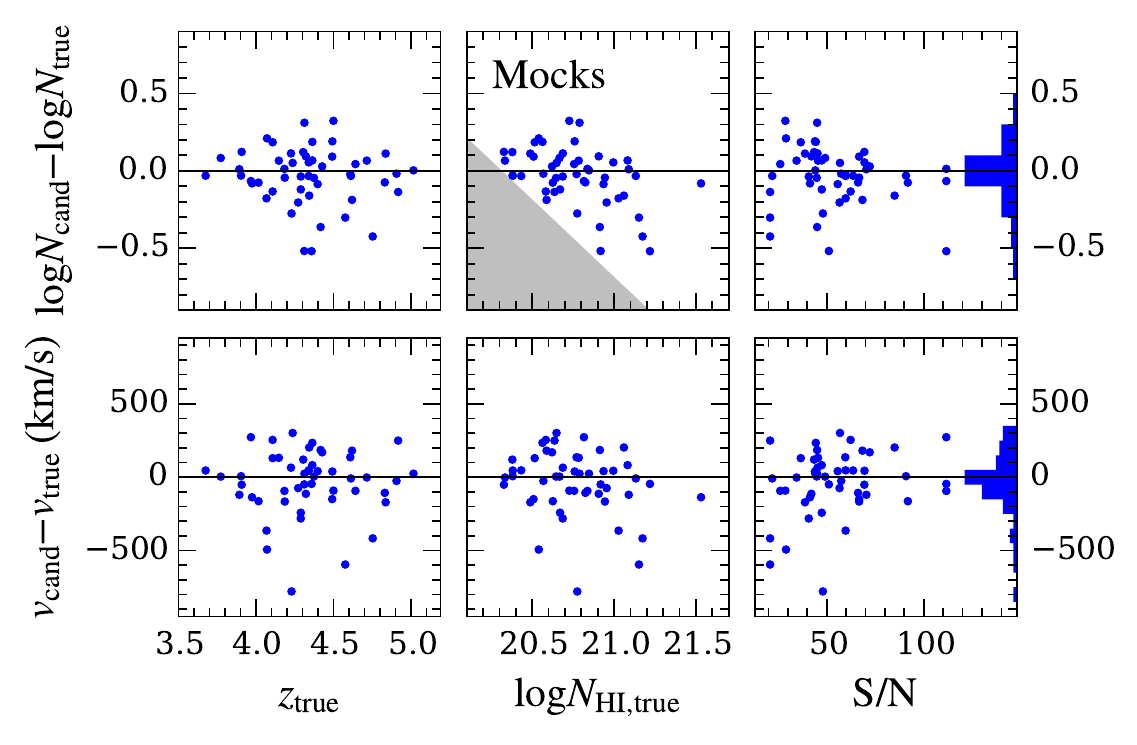}
\end{tabular}
\caption{\label{f_match} The difference between \NHI\ estimated for
  DLA candidates in low resolution spectra ($N_\mathrm{HI,cand}$) and
  $N_\mathrm{HI,true}$ measured from high resolution spectra (top) or
  known from mock linelists (bottom), and similarly for the velocity
  offset from the true DLA redshift.  This is for one of the authors
  (JXP), but the results for NHMC are similar. These show there is no
  strong systematic offset in the estimated \NHI\ as a function of
  redshift, S/N or \NHI\ which might systematically bias
  \OmHI\ significantly. Grey shading shows regions that cannot be
  populated due to the requirement that both $N_\mathrm{HI,cand}$ and
  $N_\mathrm{HI,true}$ are $>10^{20.3}$~\cmm.}
\end{figure}

The results of this test are shown in Figure~\ref{f_match}. The log
\NHI\ difference is plotted as a function of redshift, the true \NHI,
and S/N for the high-resolution sample (top panels) and mocks (bottom
panels). For both the mocks and high-resolution samples, both $\Delta
\log \NHI$ and $\Delta v$ are centred on 0. The standard deviation of
the velocity and $\log \NHI$ offsets are 184/216~\kms\ and 0.165/0.196
for the high-resolution sample and mocks, respectively.  We therefore
adopt 0.2 dex as our uncertainty in \NHI. There is no trend seen with
redshift or S/N. There may be a trend with \NHI, but above
$N=10^{20.3}~\cmm$ it is too weak to significantly affect \OmHI. We
conclude that there is no systematic bias in \NHI\ which might
adversely affect the \OmHI\ measurement.

Figure~\ref{f_match} also shows the redshift difference between
matched DLAs expressed as a velocity difference. DLAs identified in
the higher resolution spectra use low-ionization metal lines to set a
precise DLA redshift with an error a few \kms. Both the mocks and high
resolution sample show that an uncertainty of $\sim 200$~\kms\ results
from estimating redshifts using Lyman series absorption alone (without
reference to metal absorption) in the low-resolution spectra.

\subsection{DLA incidence rate and differential \NHI\ distribution}
\label{s_fN}

\begin{figure}
\begin{center}
\includegraphics[width=0.8\columnwidth]{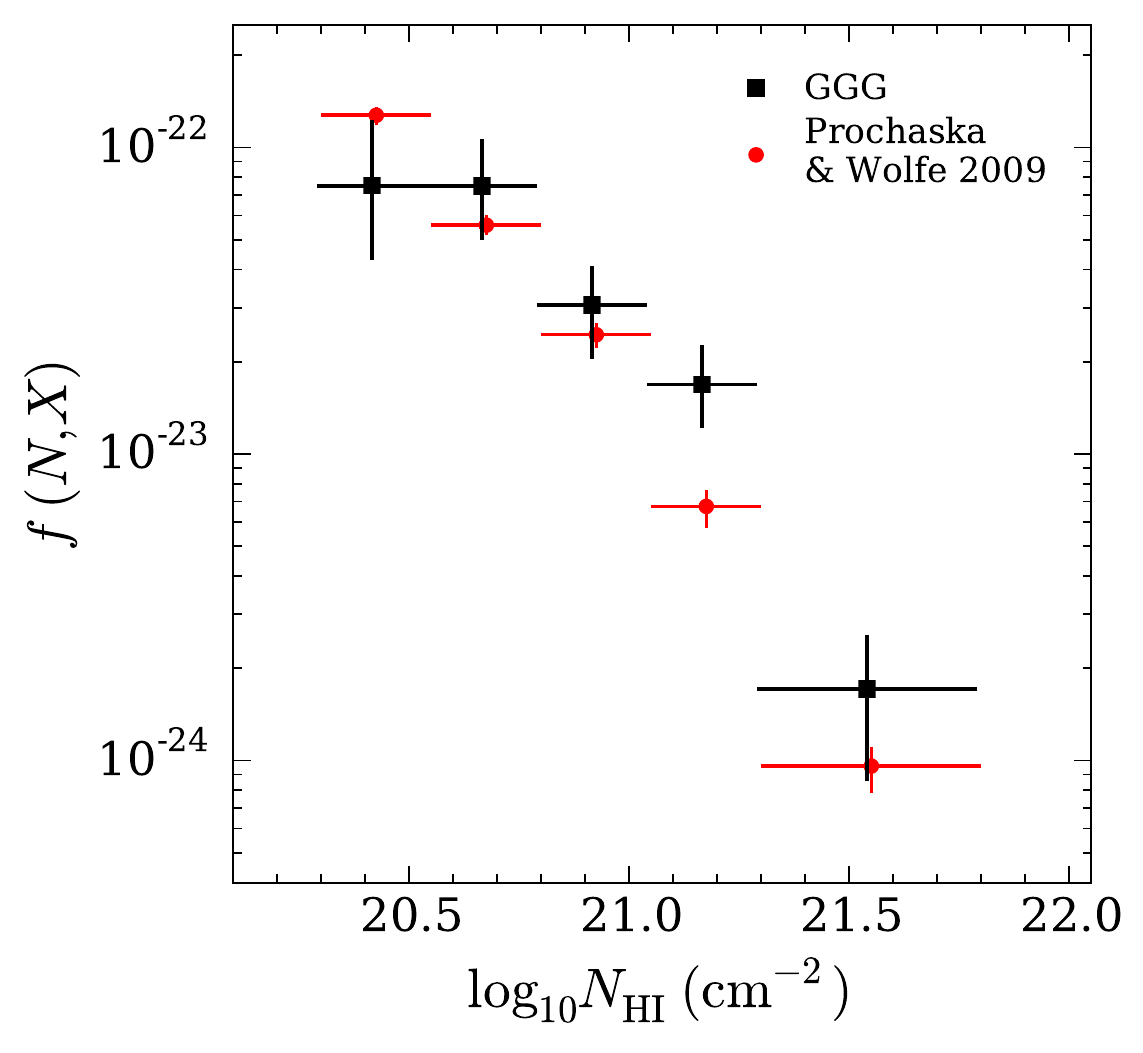}
\caption{\label{f_fNX} The column density distribution, $f(\NHI)$, for
  the GGG sample of DLAs. Black points show the measurements with
  correction $k(\NHI)$ applied. Both are slightly offset in \NHI\ for
  clarity. Red points show the SDSS DR5 measurements from
  \citet{Prochaska09_OmHI} after applying the correction to the
  redshift search path recommended by \citet{Noterdaeme09_OmHI}. The
  errors are $1\sigma$, and include statistical and systematic errors
  (see section \ref{s_fN} for more details).}
\end{center}
\end{figure}

Figure \ref{f_fNX} shows the differential \NHI\ distribution from the
GGG sample compared to that from the SDSS sample from
\citet{Prochaska09_OmHI}, which is consistent with the more recent
estimate from \citealt{Noterdaeme12_OmHI}. We have four different
measurements of the correction factor $k(\NHI)$, from two different
authors using the mocks and high-resolution spectra, so there are four
different estimates of $f(\NHI,X)$. We find the final $f(\NHI,X)$ by
averaging these four estimates. The uncertainties on this value
include a statistical and systematic component. The statistical
uncertainty is found by bootstrap resampling, using 1000 samples from
the observed DLA distribution, and averaging these uncertainties for
the four different estimates. The systematic uncertainty is then
assumed to be the standard deviation in the four estimates. These
systematic and statistical components are added in quadrature to give
the errors shown in figure \ref{f_fNX}. The two distributions are
similar overall, although there is a clear discrepancy between the GGG
and $z=3$ $f(\NHI,X)$ for the bin at $\log \NHI\sim 21.2$, which hints
at evolution in the shape of $f(\NHI,X)$ at high redshift. However, a
simple change in the normalization is also consistent with the data.

\begin{figure}
\includegraphics[width=1.0\columnwidth]{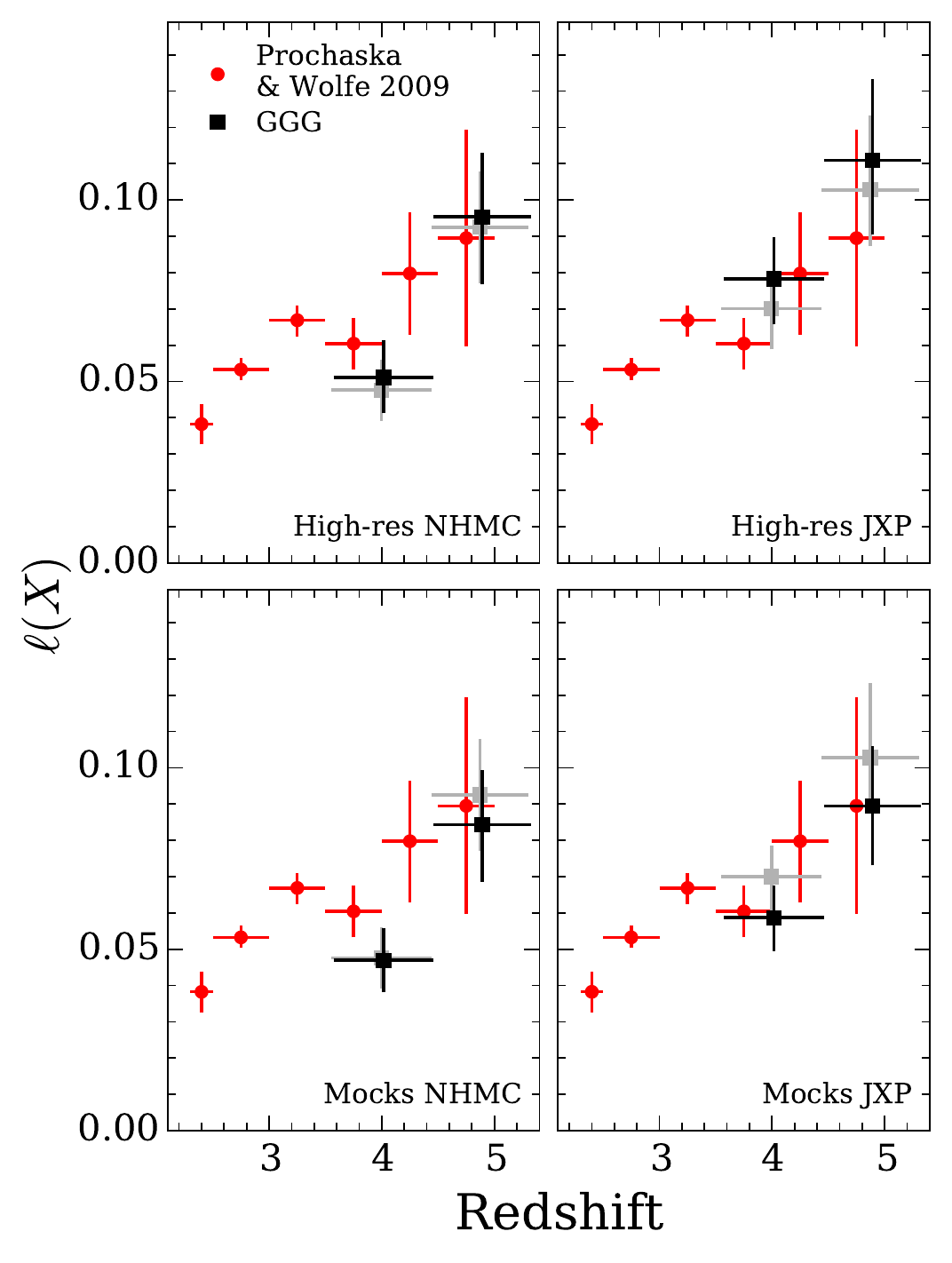}
\caption{\label{f_lX} The DLA incidence rate $\ell(X)$. This
  observable is more sensitive to the lowest \NHI\ DLAs than
  \OmHI. Grey points show the uncorrected GGG measurement, and black
  squares with the corrections applied. Each panel shows a different
  correction, using either mock or high-resolution spectra for two
  different authors. There are systematic differences comparable to
  the statistical errors, and they depend on whether the mocks or the
  high resolution spectra are used to estimate correction
  factors. Similar differences are also found between the two
  different authors who searched for DLAs. These illustrate that
  significant systematic uncertainties affect the measurement of
  $\ell(X)$.}
\end{figure}

The DLA incidence rate, $\ell(X)$, is shown in Figure~\ref{f_lX}.
This observable is more sensitive to the lowest \NHI\ DLAs than
\OmHI. Since the correction factors we derive are strongest for low
\NHI\ DLAs and these DLAs have a strong effect on $\ell(X)$, we expect
$\ell(X)$ to be sensitive to the particular choices of correction
factors. This is indeed the case -- there are systematic differences
at least as large as the statistical errors, and they depend on
whether the mocks or the high resolution spectra are used to estimate
the correction factor. Similarly large differences are found between
$\ell(X)$ by each of the two authors who searched for DLAs. The
$\ell(X)$ values we measure are consistent with a smooth increase from
$z=2$ to $z=5$. However, since we do not know which $k(\NHI)$
correction factors are best, we do not attempt to present a definitive
$\ell(X)$ measurement here. A large sample of higher-resolution
spectra, where low column density DLAs can be identified with more
certainty, will be necessary to robustly measure $\ell(X)$ at $z>4$.

\begin{figure}
\begin{center}
\includegraphics[width=0.8\columnwidth]{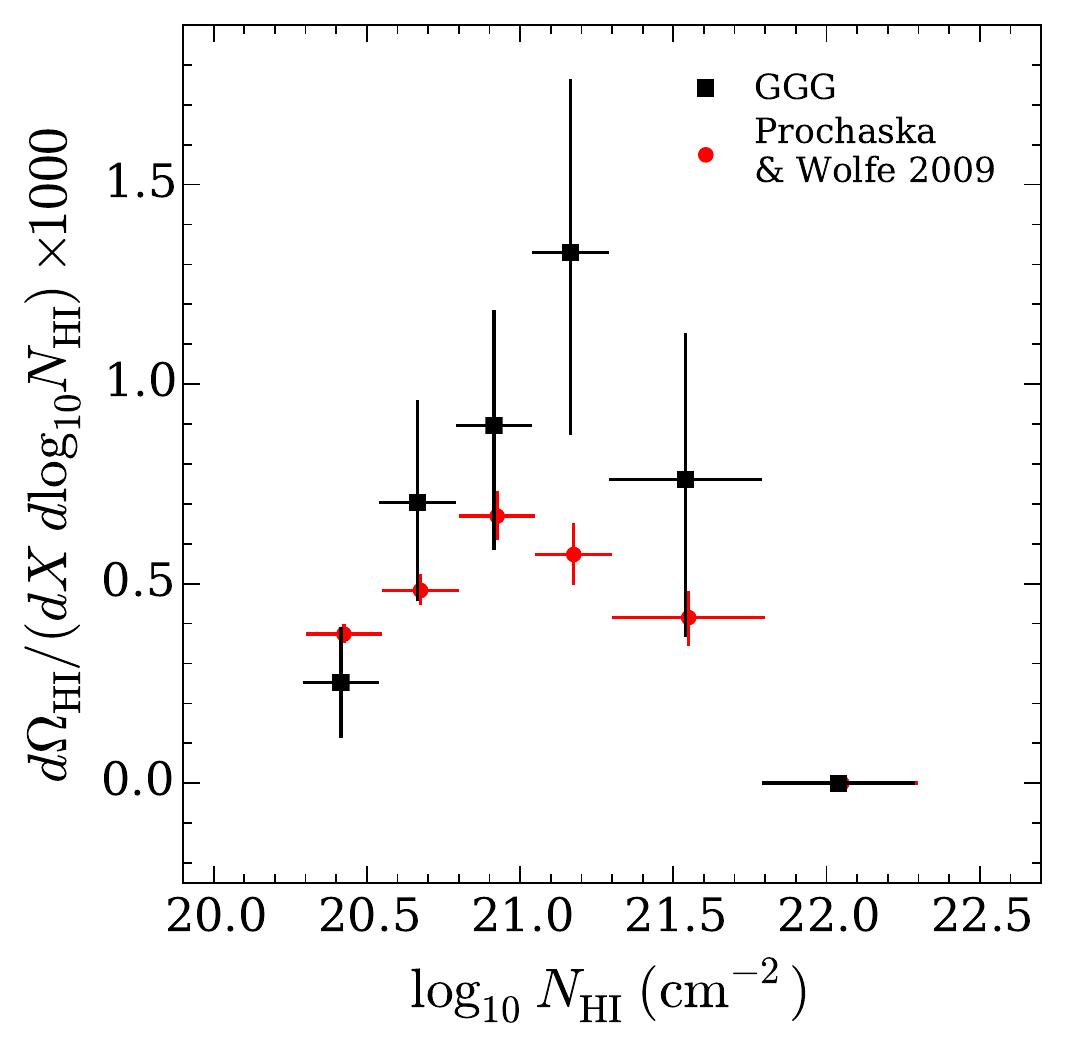}
\caption{\label{f_dOX} The differential \OmHI\ distribution,
  $d\OmHI/(dXd\log \NHI)$. DLAs with the largest contribution to
  \OmHI\ have \NHI\ in the range $10^{20.8}$--$10^{21.6}$~\cmm; DLAs
  with lower \NHI\ are less important. Therefore, while there is
  uncertainty in the number of low column density systems (and thus
  $\ell(X)$), \OmHI\ can still be measured accurately. This is also
  illustrated by Figure~\ref{f_Om_sys}.}
\end{center}
\end{figure}

We can still make a more robust measurement of \OmHI, however, regardless
of the uncertainty in $\ell(X)$, as Figure~\ref{f_dOX}
illustrates. DLAs with the largest contribution to \OmHI\ have
\NHI\ in the range $10^{20.8}$--$10^{21.6}$~\cmm, and DLAs with lower
\NHI\ make a substantially smaller contribution. Therefore, while
systematic effects may give rise to a large uncertainty in the number
of low column density systems (and thus $\ell(X)$), \OmHI\ can still
be measured accurately. This point is discussed further in
Section~\ref{s_OmHI}.


\section{Results and discussion}
\label{s_OmHI}
\subsection{\OmHI\ measurement}
\label{s_OmHI_meas}
We can now use the \NHI-dependent correction factor $k$ estimated in
the previous section to find $f_\mathrm{DLA}$ and thus \OmHI. For the
GGG sample we count the number of DLAs in a given absorption path,
giving each DLA a weight $k(\NHI)$, where
$k(\NHI)=k_\mathrm{found}(\NHI)/k_\mathrm{real}(\NHI)$. $k(\NHI)$ is
then estimated as the ratio of the $\log \NHI$ histograms shown in
Figures \ref{f_kfound} and \ref{f_kreal}, with the uncertainty on each
bin given by the uncertainties in $k_\mathrm{found}$ and
$k_\mathrm{real}$ added in quadrature.

There are two main contributions to the final error on \OmHI. The
dominant contribution is the statistical error due to the finite
sampling of DLAs: there are $25$--$30$ DLA candidates in each redshift
bin, dependent on whether NHMC or JXP's results are used.  We estimate
this error using 1000 bootstrap samples from the DLA sample. The
second is the systematic uncertainty in the correction factor,
$k(\NHI)$. We estimate the effect of this uncertainty using a Monte
Carlo technique. \OmHI\ is calculated 1000 times, each time drawing
$k(\NHI)$ from a normal distribution with a mean given by the
$k(\NHI)$ histogram bin value and $\sigma$ determined by the
uncertainty on that bin, assuming no correlation between uncertainties
in adjacent bins. Then the final error in \OmHI\ is given by adding
these two uncertainties in quadrature.  We confirmed that \NHI\ error
of each DLA (0.2 dex, see Section~\ref{s_N_z}), has a negligible
contribution compared to these statistical and systematic
uncertainties. We also check that using \NHI\ measurements from the
high-resolution spectra, where available, does not significantly
change \OmHI.

\begin{figure}
\includegraphics[width=1.0\columnwidth]{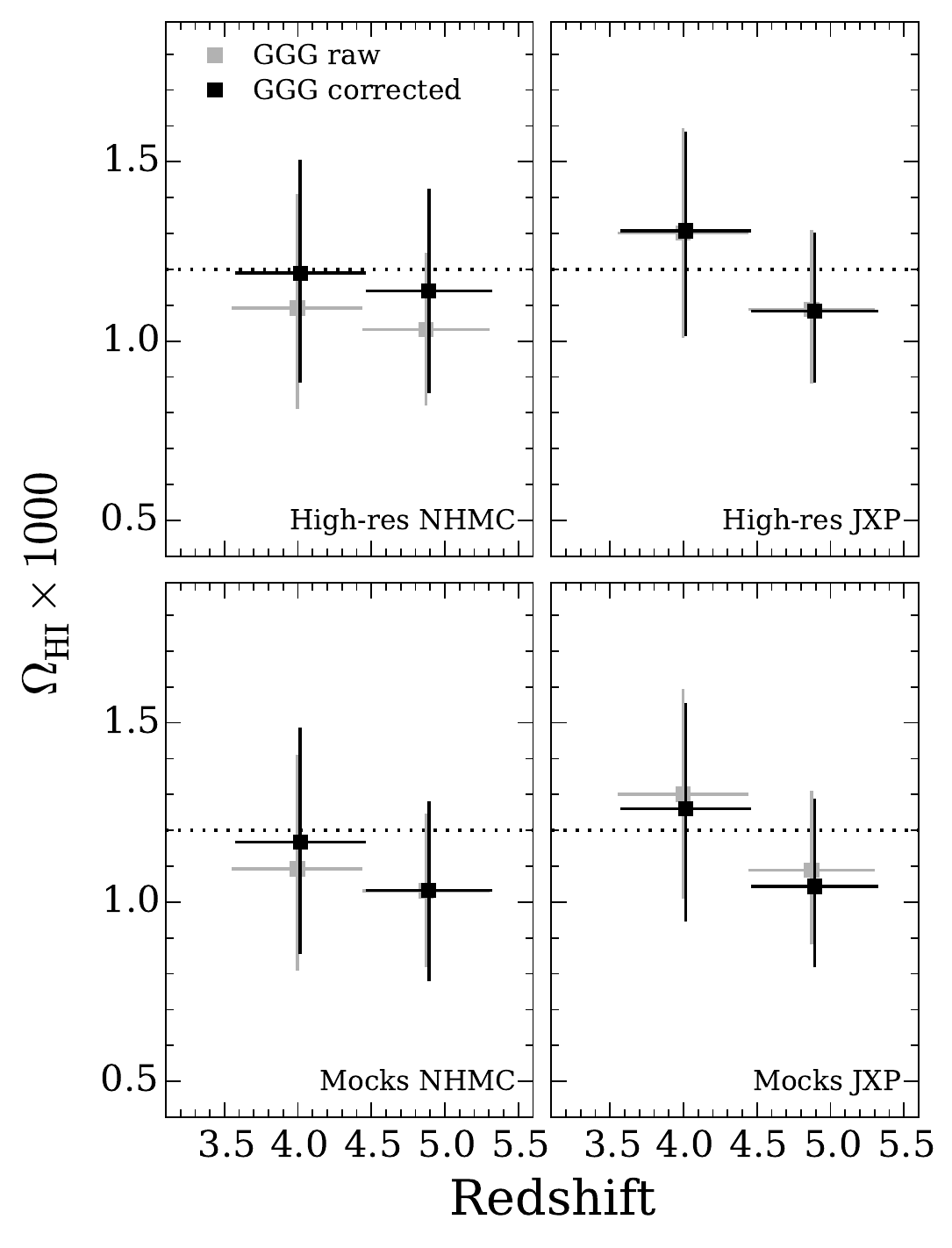}
\caption{\label{f_Om_sys} \OmHI\ measured by the two authors using the
  high-resolution sample (top) and mocks (bottom). The uncertainties
  on \OmHI\ introduced by any differences in selecting DLA candidates
  between the authors, or between using mocks or the high-resolution
  sample, are much smaller than the errors shown, which are a
  combination of the statistical error and uncertainty in the
  correction factor (see section \ref{s_OmHI}).}
\end{figure}

Since we have separate estimates of $k(\NHI)$ from the mocks and high
resolution sample, and two authors performed these estimates, we can
make 4 different measurements of \OmHI. We use these to gauge the
effect on \OmHI\ of estimating corrections from the mocks versus the
high-resolution sample, or of any differences in the way the two
authors identified DLAs. The results are shown in Figure
\ref{f_Om_sys}. The differences between the mocks compared to the
high-resolution sample, and between the two authors, are significantly
smaller than the uncertainty on any individual
\OmHI\ measurement. Therefore we conclude that neither the methods we
use to estimate $k(\NHI)$, nor any differences in DLA detection
between methods, contribute a significant uncertainty to the final
\OmHI. We caution that this conclusion only holds for the sample of
spectra we analyse. New tests of systematic effects may be required
for measurements of \OmHI\ using larger samples of DLAs, or using
different resolution or S/N QSO spectra.

For the remainder of the paper we use the measurement of
\OmHI\ derived using $k$ from the higher-resolution sample and
measured by author JXP, which is shown in the top-right panel of
Figure~\ref{f_Om_sys}. This measurement and the 68\% confidence
interval is given in Table~\ref{t_OmHI}. We assume a 20\% contribution
to \OmHI\ from systems below the DLA threshold, as described in
Section\ref{s_deltaHI}.

\begin{table}
\addtolength{\tabcolsep}{4pt}
\begin{center}
\begin{tabular}{cccc}
\hline
$ z $ & $10^3\, \OmHI$ & $10^3\, \OmHI$ $(1\sigma)$ & $\Delta X$ \\
\hline
3.56-4.45 & 1.18 & 0.92-1.44 & 356.9 \\
4.45-5.31 & 0.98 & 0.80-1.18 & 194.6 \\
\hline
\end{tabular}
\caption{\label{t_OmHI} \OmHI\ for the GGG sample, assuming a flat
  cosmology with $H_0=70$~\kmsMpc\ and $\Omega_{m, 0} = 0.3$. The
  redshift bins were chosen to cover roughly equal redshift widths,
  and to yield approximately equal numbers of DLAs in each bin. To
  convert between \OmHI\ and \OmDLAg, which is often quoted by other
  DLA studies, use $\OmHI=\delta_{\rm HI}\OmDLAg/\mu$, where $\mu =
  1.3$ accounts for the mass of helium and $\delta_{\rm HI}=1.2$
  estimates the contribution from systems below the DLA threshold of
  $10^{20.3}$~\cmm.}
\end{center}
\end{table}

\subsubsection{Is there a bias from gravitational lensing?}
\label{s_is_lensing}

There is a $30\pm20$\% increase in \OmHI\ for sightlines towards the
brighter half of our QSO sample ($z \le 19.2$~mag) relative to
\OmHI\ towards the fainter QSOs ($z>19.2$~mag). If this effect is
caused by gravitational lensing of a background QSO by a galaxy
associated with a foreground DLA, then our measured \OmHI\ will be
artificially enhanced.  A detailed lensing analysis is beyond the
scope of this work. However, if we follow \citet{Menard12} and assume
the lensing DLA galaxies are isothermal spheres, we can estimate their
Einstein radius as
\begin{equation}
\zeta_0 = 4 \pi \left( \frac{\sigma_v}{c} \right)^2 \frac{D_l
  D_{ls}}{D_s}
\end{equation}
where $\sigma_v$ is the velocity dispersion, $c$ is the speed of light
and $D_{l,s,ls}$ are the angular diameter distances from the observer
to the lens and to the source, and from the lens to the source.
Assuming a typical dispersion of $100$~\kms\ we find the effective
radius for lensing is very small, $0.1$~kpc for a $z=4.5$ DLA towards
a $z=5$ QSO. This is half the radius for a DLA at $z=2.5$ towards a
QSO at $z=3.5$. Since the magnitude of the increase in \OmHI\ due to
the putative lensing at $z\sim3$ is relatively small ($\sim 20\pm10$
per cent, Prochaska et al. 2005) we do not expect it to have a large
effect at higher redshifts. We conclude that it is more likely the
difference in \OmHI\ between the bright and faint QSO samples is
caused by a statistical fluctuation, rather than a lensing bias.

\subsection{Comparison with previous measurements}

\label{s_prev}

Several groups have made measurements of \OmHI\ at $z>4.5$ using DLA
surveys (\citealt{Peroux03}, \citealt{Guimaraes09}, S10).  These are
cumulative results -- \OmHI\ measurements from each new QSO sample are
combined with older \OmHI\ measurements which used a different DLA
survey. While combining results in this way maximizes the statistical
S/N of the final result, it results in a heterogeneous sample of
quasar spectra with different data quality and different DLA
identification methods. As shown in sections \ref{s_method} and
\ref{s_OmHI_meas}, at $z>4.4$ different identification methods can
produce a systematic uncertainty in \OmHI\ which, although smaller
than the statistical uncertainties for our current DLA sample, may
still be considerable. Since these analyses did not use mock spectra
to explore systematic effects, it is difficult to estimate the true
uncertainty in \OmHI\ when combining heterogeneous quasar samples with
different selection criteria. In contrast, our sample has homogeneous
data quality, QSO selection method and DLA identification procedure,
and we use mock spectra to test any systematic effects.\footnote{We
  note that eight of the QSOs used by S10 are also included in our
  sample, but the 155 remaining GGG QSOs are independent of previous
  samples.}

\begin{figure*}
\begin{tabular}{c}
\includegraphics[width=0.65\textwidth]{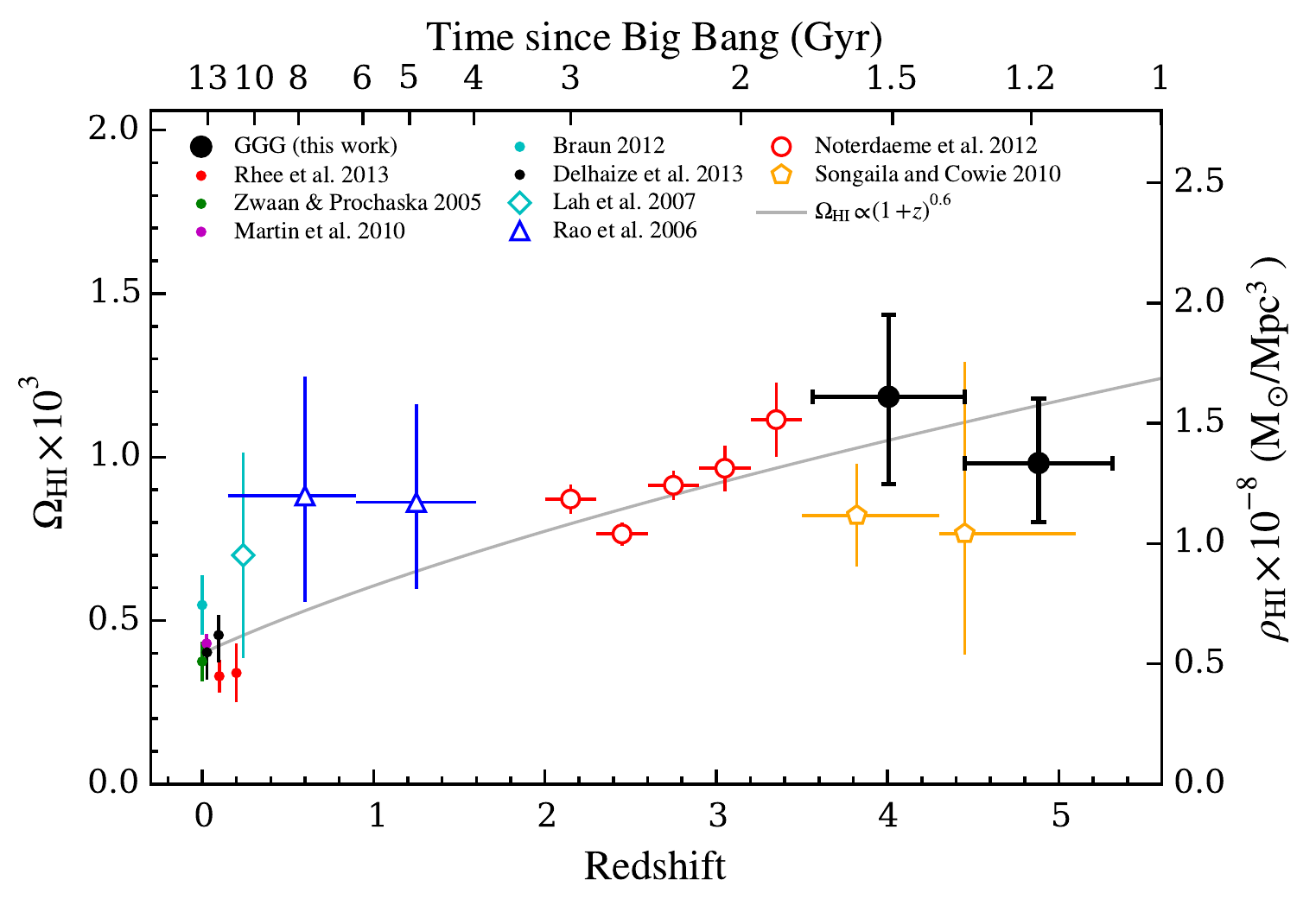}
\end{tabular}
\caption{\label{f_Om_all} Measurements of \OmHI\ at different
  redshift, from
  \citet{Zwaan05,Rao06,Lah07,Braun12,Martin10_OmHI,Noterdaeme12_OmHI,Rhee13,
    Delhaize13} and S10 (see Table \ref{t_literature}). We do not show
  the measurement using SDSS QSOs by \citet{Prochaska09_OmHI}, it is
  consistent with the measurement by \citet{Noterdaeme12_OmHI}, who
  use a superset of SDSS QSOs.  We also do not show the
  \citet{Peroux03} and \citet{Guimaraes09} results, which have a large
  overlap with the QSO sample used by S10 and are consistent with that
  measurement. Finally, for clarity we do not show the measurements at
  lower redshift from \citet{Freudling11} and \citet{Meiring11_OmHI};
  they are consistent with the plotted values.  All measurements have
  been converted to the same cosmology ($h=0.7$, $\Omega_{m}=0.3$,
  $\Omega_{\Lambda}=0.7$) and include \HI\ mass only, with no
  contribution from helium or molecular hydrogen.}
\end{figure*}

Figure~\ref{f_Om_all} shows our new results together with previous
measurements of \OmHI, converted to our adopted cosmology. When
multiple measurements of \OmHI\ have been made using overlapping QSO
samples and the most recent measurement uses a superset of previous
QSO samples, only the most recent measurement is shown.  For example,
the results of S10 include most of the quasars used by
\citet{Peroux03} and \citet{Guimaraes09}, so we show only the S10
result. In all such cases the most recent measurement is consistent
with earlier results. Where previous DLA surveys have quoted $\OmDLA$,
we convert to $\OmHI$ using the relationship $\OmHI = 1.2 \OmDLA /
1.3$.  Our measurement at $\langle z\rangle=4$ is higher than, but
consistent with earlier measurements by S10. As such and because we
find a possible systematic increase in \OmHI\ towards bright QSOs, we
checked whether the magnitude distribution of the S10 QSOs was lower
than the GGG sample. $z$ band data was not available for the whole S10
sample, but the eight QSOs which overlap between their sample and ours
have a similar fraction of QSOs with $z\le19.2$ and
$z>19.2$~mag. Therefore a difference in QSO magnitudes is unlikely to
cause a difference between our result and the S10 result, and it seems
more likely that the difference is caused by a statistical
fluctuation.

Our results at $\langle z\rangle=4.9$ give the most robust indication
to date that there is no strong evolution in \OmHI\ over the
$\sim1$~Gyr period from $z=5$ to $z=3$. We see a slight drop in
\OmHI\ between our $z\sim4$ and $z\sim4.9$ \OmHI\ measurements, but
this difference is not statistically significant. If the metal content
of DLAs does change suddenly at $z=4.7$, as suggested by
\citet{Rafelski14}, there is no evidence it is accompanied by a
concomitant change in \OmHI. However, the uncertainties remain large
and future observations should continue to test this possibility.

Figure~\ref{f_Om_all} also shows a power law with the form
$\Omega_{\rm HI} = A(1+z)^\gamma$ fitted to the binned data. This simple
function provides a reasonable fit ($\chi^2$ per degree of freedom
$=1.44$) across the full redshift range, with best-fitting parameters
$A=(4.00\pm0.24)\times 10^{-4}$ and $\gamma=0.60\pm0.05$. There is no
obvious physical motivation for this relation, nor any expectation
that it should apply at redshifts $>5$. Nevertheless, it may provide a
useful fiducial model to compare to simulations and future
observations.

\begin{figure}
\begin{center}
\includegraphics[width=0.85\columnwidth]{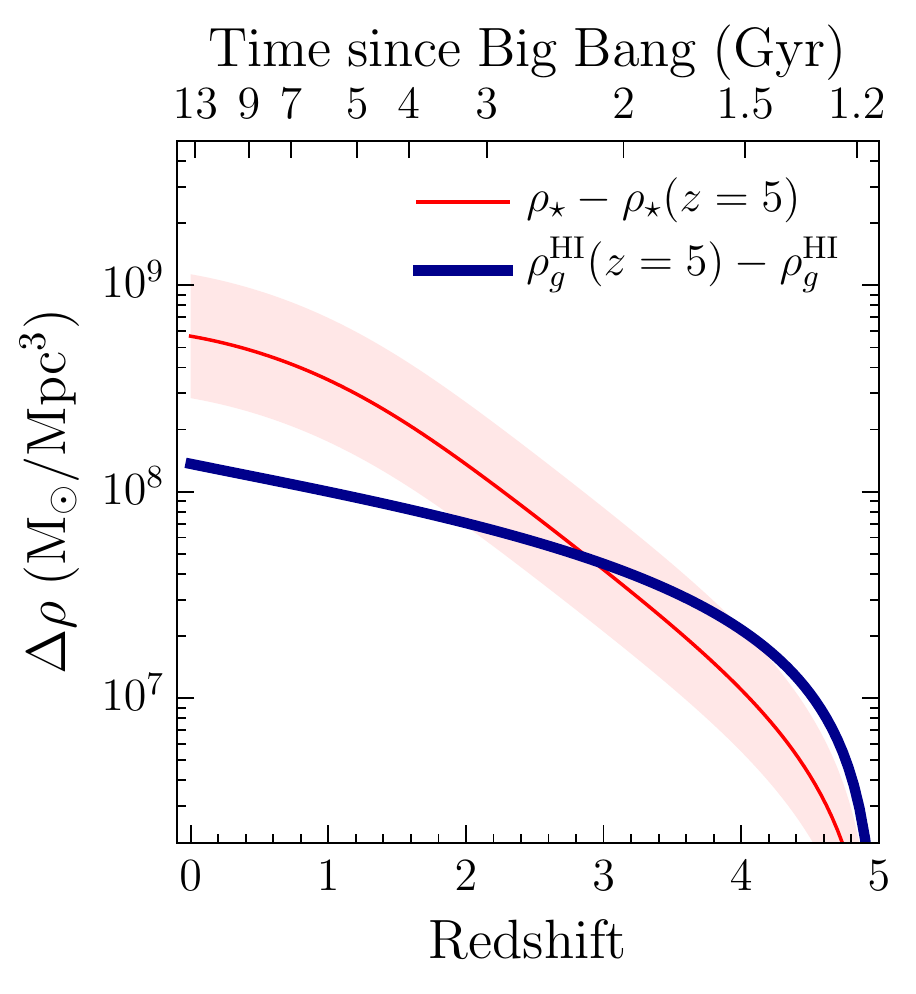}
\caption{\label{f_rho_ev} The increase in comoving stellar mass
  density from $z=5$ to $0$ \citep[from][thin line and
    shading]{Madau14} and the corresponding decrease in \HI\ gas mass
  density over the same period (thick line) using the fitting formula
  from section \ref{s_prev}. Before $z\sim3$, the \HI\ gas phase
  contains ample mass density to fuel all the observed star
  formation. However, from $z\sim3$ to the present it contributes less
  than $\sim20\%$ of the mass necessary to form stars, and so must be
  continually replenished by more highly ionized gas.}
\end{center}
\end{figure}

We also compare our new high-redshift value to lower redshift
\OmHI\ measurements. As previous authors have noted
\citep[e.g.][]{Prochaska05_DLA, Prochaska09_OmHI, Noterdaeme09_OmHI},
\OmHI\ evolves from $z=3$ to $z=0$ by factor of $\lesssim2$, at odds
with the very strong evolution in the star formation rate over the
same period. Moreover, the drop in \OmHI\ is much smaller than the
increase in stellar mass over this period. Figure \ref{f_rho_ev}
demonstrates this point by showing the increase in comoving mass
density in stars from $z=5$, $\rho_\star-\rho_\star(z=5)$ and the
contemporaneous decrease in \HI\ comoving gas mass density\footnote{In
  Figure \ref{f_rho_ev} the \HI\ gas mass density $\rho^\mathrm{HI}_g$
  is used, which is related to the \HI\ mass density by
  $\rho^\mathrm{HI}_g \equiv \mu \rho_\mathrm{HI}$ with $\mu = 1.3$
  and $\rho_\mathrm{HI}=\rho_{\rm crit,0}\OmHI$. We do not apply any
  correction for dust extinction by foreground DLAs. If this is
  present, it could increase $\rho^\mathrm{HI}_g$ by $20\%$
  \citep{Pontzen09}, which would not affect our discussion.},
$\rho^\mathrm{HI}_g(z=5)-\rho^\mathrm{HI}_g$ using the power law fit
from Figure \ref{f_Om_all}.  The mass in stars is calculated using the
expression from \citet{Madau14}, and the range shows an uncertainty of
$50\%$, indicative of the scatter in observations around this
curve. While the evolution of \OmHI\ from $z=5$ to $z=3$ remains
uncertain, the \HI\ phase at $z=5$ contains ample mass density to form
all the stars observed at $z\sim3$, and the evolution predicted by the
simple power law function is consistent with this scenario. From
$z\sim3$ to $z\sim0$, however, there is a factor of $5$--$6$ shortfall
in \HI\ mass density compared to amount needed to produce stars over
the same period. This underscores that at $z\lesssim3$, the \HI\ phase
must be continually replenished by more highly ionized gas, presumably
through a combination of cold-mode accretion \citep[e.g.][]{Dekel09}
and recycled winds \citep[e.g.][]{Oppenheimer10}. The more highly
ionized Lyman limit systems and sub-DLAs should then be important
tracers of the interface between this \HI\ phase and more highly
ionized gas \citep[e.g.][]{Fumagalli11}.

There are several reasons to expect the neutral fraction of the
universe to evolve at $z>3$. As we approach the epoch of reionization,
the filling factor of neutral hydrogen in the universe should
increase, as large pockets of the universe are no longer ionized. This
is reflected in the decrease in the mean free path for H-ionizing
photons \citep{Fumagalli13,Worseck14_GGG} towards higher
redshifts. While the bulk of reionization is thought to occur at
$z>6$, large neutral regions may persist to lower redshifts
\citep[e.g.][]{Becker15}. Our results suggest that while regions of
this kind may exist, they do not change the total neutral gas mass
density appreciably from that observed at $z\sim3$. This is consistent
with the conclusions of Becker et al., who find that by $z=5$ the bulk
of IGM absorption is due to density fluctuations instead of large,
neutral regions yet to be reionized.

This is perhaps not surprising. The distribution of these neutral
pockets depends on the nature of reionization, which may progress from
low-density regions to high-density regions (`outside--in') or the
reverse (`inside--out'), or some combination of the two
\citep[e.g.][]{Finlator12}.  However, favoured scenarios see the
highest density regions with
$\Delta\equiv\rho/\langle\rho\rangle\gg100$ reionized first, as they
are populated by galaxies, believed to be the dominant source of
ionizing photons. In this case neutral pockets will persist only in
underdense regions such as filaments or voids, with $\Delta<100$. At
$z\sim 2.5$ clustering measurements suggest most DLAs are found inside
haloes with masses $10^{10}$--$10^{12}$~\msun
\citep{Cooke06,FontRibera12}, which have a mean
$\Delta>100$. Therefore even if large neutral regions do persist to
$z=5$, they may not occur at cosmic densities high enough to produce
strong DLA absorption. The remnants of such regions may be observable
as Lyman limit systems however, and so one might expect an increase in
their incidence rate towards $z\sim5$, which observations already hint
may be the case \citep{Prochaska10,Fumagalli13}. The GGG sample can
also be used to measure the LLS incidence rate at $z>4$, which we will
present in a future work.

\begin{table}
\addtolength{\tabcolsep}{4pt}
\begin{center}
\begin{tabular}{ccc}
\hline
$z$       & $10^3\, \OmHI$         &  Reference                \\
\hline
0         & $0.375 \pm 0.061$         &  Zwaan et al. (2005)	  \\
0         & $0.548 \pm 0.091$         &  Braun (2012)     	  \\
0.026     & $0.430 \pm 0.030$         &  Martin et al. (2010)	  \\
0.028     & $0.403^{+0.043}_{-0.084}$ &  Delhaize et al. (2013)	  \\
0.096     & $0.456^{+0.061}_{-0.084}$ &  Delhaize et al. (2013)	  \\
0.1       & $0.33  \pm 0.05$          &  Rhee et al. (2013)       \\
0.2       & $0.34  \pm 0.09$          &  Rhee et al. (2013)       \\
0.24      & $0.70  \pm 0.31$          &  Lah et al. (2007)       \\
0.15-0.90 & $0.88^{+0.36}_{-0.33}$    &  Rao et al. (2006)        \\
0.9-1.6   & $0.86^{+0.30}_{-0.27}$    &  Rao et al. (2006)        \\
2.0-2.3   & $0.872 \pm 0.044$         &  Noterdaeme et al. (2012) \\
2.3-2.6   & $0.765 \pm 0.035$         &  Noterdaeme et al. (2012) \\
2.6-2.9   & $0.914 \pm 0.044$         &  Noterdaeme et al. (2012) \\
2.9-3.2   & $0.966 \pm 0.070$         &  Noterdaeme et al. (2012) \\
3.2-3.5   & $1.11  \pm 0.11$          &  Noterdaeme et al. (2012) \\
3.5-4.3   & $0.82^{+0.30}_{-0.27}$    &  Songaila (2010)          \\
4.3-5.1   & $0.77^{+0.30}_{-0.27}$    &  Songaila (2010)  \\
\hline
\end{tabular}
\caption{\label{t_literature} \OmHI\ measurements from the literature
  shown in Figure \ref{f_Om_all}. Each has been converted to a flat
  cosmology with $H_0=70$~\kmsMpc\ and $\Omega_{m, 0} = 0.3$, and
  represent the mass density from \HI\ gas alone, without any
  contribution from helium or molecules. For previous analyses which
  quote the gas mass in DLAs, \OmDLAg, we have converted to
  \OmHI\ using $\OmHI=\delta_{\rm HI}\OmDLAg/\mu$, where $\mu = 1.3$
  accounts for the mass of helium and $\delta_{\rm HI}=1.2$ estimates
  the contribution from systems below the DLA threshold of
  $10^{20.3}$~\cmm\ (see section \ref{s_deltaHI}).}
\end{center}
\end{table}

\subsection{Comparison with theory}

\begin{figure*}
\begin{tabular}{c}
\includegraphics[width=0.65\textwidth]{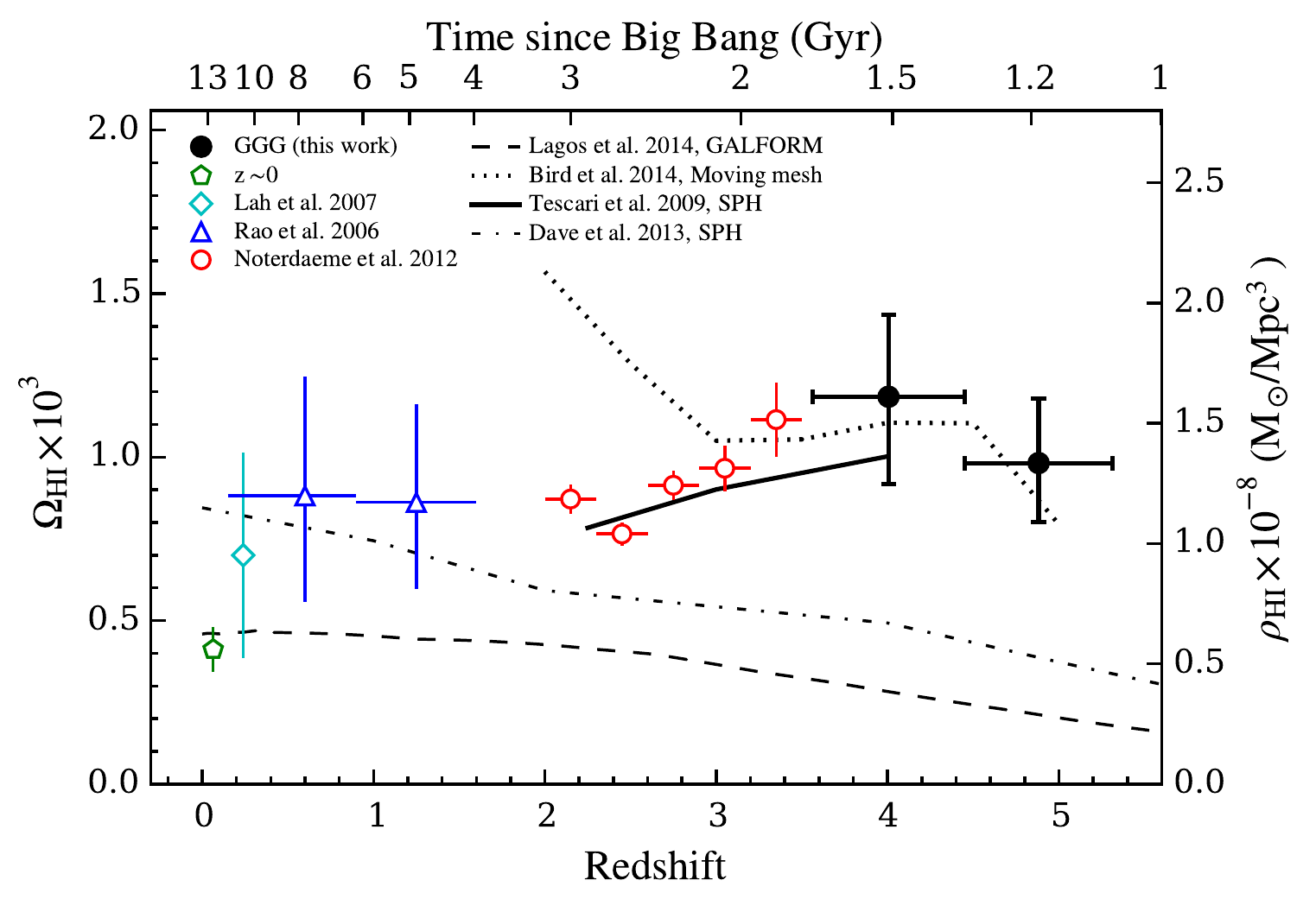}
\end{tabular}
\caption{\label{f_Om_theory} Measurements of \OmHI\ compared to recent
  theoretical predictions. For clarity, the mean of measurements at $z
  < 0.2$ (the errorbar shows the standard deviation) is shown. Lines
  show predictions from a recent semi-analytic model \citep{Lagos14},
  along with SPH \citep{Tescari09,Dave13} and moving-mesh
  \citep{Bird14_DLAs} simulations. All the models have been converted
  to our adopted cosmology. While they do reproduce the roughly flat
  evolution of \OmHI\ from $z=5$ to $0$ (in comparison to the cosmic
  star formation rate density), they do not match the data across the
  full redshift range.}
\end{figure*}

In Figure \ref{f_Om_theory} we show \OmHI\ in comparison to some
recent theoretical predictions for its evolution.  These are by
\citet{Lagos14} using the semianalytic GALFORM model, by
\citet{Bird14_DLAs} from a simulation using the moving-mesh code
AREPO, and by Tescari et al. (\citeyear{Tescari09}, see also
\citealt{Duffy12_OmHI}) and \citet{Dave13} using smoothed particle
hydrodynamics (SPH) simulations.  While these models broadly match the
slow evolution of \OmHI\ since $z\sim4$, most struggle to reproduce
the trend of decreasing \OmHI\ with time (with Tescari et al. being a
notable exception). Lagos et al. suggest that their model's
underestimation of \OmHI\ at high redshift may be due to more neutral
gas being found outside galaxy discs in the early universe.  If this
interpretation is correct, then our observations suggest that more
than half the neutral gas mass (and more than half of DLAs) are found
outside galaxies at $z\sim5$. Alternatively, \citet{Dave13} show that
agreement between their simulations and observations can be improved
by assuming that a population of low mass galaxies, unresolved by
current SPH simulations, make a significant contribution to the DLA
absorption cross-section at high redshift.

It is evident that further improvements are needed to theoretical
models to reproduce the evolution of \OmHI\ across the full redshift
range. If much of the neutral gas is found in galactic outflows or
recycled winds, the sub-grid prescription for outflows in SPH
simulations may have a strong influence on the predicted
\OmHI\ \citep[e.g.][]{Bird14_DLAs}. Furthermore, given the small sizes
of DLAs ($\sim5$~kpc, \citealt{Cooke10}) it may also be important to
correct for any smoothing over small-scale density peaks where DLAs
are produced, and account for hydrodynamic instabilities which are not
resolved by current cosmological simulations
\citep[e.g.][]{Crighton15_LAE}.

\section{Summary}
\label{s_summary}

We have measured \OmHI\ at $3.5<z<5.3$ using the Giant Gemini GMOS
Survey, a homogeneous sample of 163 QSO spectra with emission
redshifts $>4.4$. All the QSOs were colour-selected from the SDSS
survey and so have a well-understood selection function which is
independent of any strong absorption in the QSO spectra. Using a
combination of higher-resolution spectra of DLA candidates and mock
spectra, we explore systematic uncertainties in identifying DLAs due
to strong IGM absorption at high redshift and the low spectral
resolution of the GMOS spectra. The main conclusions from our analysis
are:
\begin{itemize}
\item We derive the most precise measurement of \OmHI\ at $\langle
  z\rangle=4.9$ to date, with a redshift path length at $z>4.5$ a
  factor of eight larger than previous analyses.  \OmHI\ at $z=4.5$ is
  consistent with the value measured at $z=3$--$3.5$, and there is no
  evidence that \OmHI\ evolves strongly over the Gyr period from
  redshift 5 to 3. There is also no evidence for an abrupt change in
  \OmHI\ between $z=4$ and $z=5$, which may be associated with a
  sudden change in metallicity reported at a similar redshift
  \citep{Rafelski14}. However, such a change is not strictly ruled out
  by the data.
\item We quantify and correct for the fraction of spurious DLA
  candidates, and for any DLAs missed in the low-resolution spectra,
  using higher resolution and mock spectra. We also estimate the
  uncertainty in the DLA column densities. For this DLA sample, the
  uncertainty introduced by these systematic effects on the
  \OmHI\ measurement is smaller than the statistical uncertainties.
\item Using the higher resolution spectra and mocks we show that the
  typical uncertainty on the DLA \NHI\ and redshift is 0.2 dex and
  200~\kms, respectively. Despite the increased IGM absorption at
  higher redshifts and the low spectral resolution, we find no strong
  systematic offset in the estimated \NHI\ for DLAs either as a
  function of redshift, or \NHI.
\item We find an excess in \OmHI\ ($30\pm20$ per cent) from the
  brighter half of our QSO sample compared to the fainter half. This
  is consistent with similar effects found in previous analyses at
  $z\sim2.5$, which posited gravitational lensing as a possible
  explanation. Given the smaller Einstein radius at $z=4.5$ compared
  to $z=2.5$, for our sample this effect seems more likely to be
  caused by a statistical fluctuation. As such it should not
  significantly bias our result.
\item Recent theoretical models do not match the data across their
  full redshift range ($z=5$ to $0$). A simple power law model of the
  form $\Omega_\mathrm{HI} = A(1+z)^\gamma$ with
  $A=(4.00\pm0.24)\times 10^{-4}$ and $\gamma=0.60\pm0.05$, while not
  physically motivated, does describe the observations over the entire
  redshift range.
\end{itemize}

\section*{Acknowledgments}

We thank Marcel Neeleman for comments on an earlier version of this
paper, Regina Jorgenson for providing a MagE spectrum of one of the
GGG DLAs, and the referee for their suggestions. Simeon Bird, Claudia
Lagos, Romeel Dav\'e and Edoardo Tescari kindly provided tables of
their model predictions. We thank the late Arthur M. Wolfe for
providing unpublished ESI spectra and for early contributions to this
work.

Our analysis made use of \textsc{astropy} \citep{Astropy13},
\textsc{xidl}\footnote{\url{http://www.ucolick.org/~xavier/IDL}} and
\textsc{matplotlib} \citep{Hunter07}.  NC and MM thank the Australian
Research Council for \textsl{Discovery Project} grant DP130100568
which supported this work. MF acknowledges support by the Science and
Technology Facilities Council [grant number ST/L0075X/1]. SL has been
supported by FONDECYT grant number 1140838 and received partial
support from Center of Excellence in Astrophysics and Associated
Technologies (PFB 06).

Based on observations obtained at the Gemini and W. M. Keck
Observatories. The authors wish to acknowledge the very significant
cultural role and reverence that the summit of Mauna Kea has always
had within the indigenous Hawaiian community.  We are most fortunate
to have the opportunity to conduct observations from this mountain.

\footnotesize{

}

\clearpage

\begin{table}
\renewcommand{\arraystretch}{1}
\addtolength{\tabcolsep}{-2pt}
\begin{center}
\begin{tabular}{ccccccc}
\hline
QSO name & $z_\mathrm{em}$ & Origin & S/N & GGG? \\
\hline
SDSS~J000749.16$+$004119.6& 4.78 & ESI  &   9.7 & no\\  
SDSS~J001115.23$+$144601.8& 4.97 & MAGE &  31.4 & yes\\ 
SDSS~J005421.42$-$010921.6& 5.02 & ESI  &  16.7 & no \\ 
SDSS~J021043.16$-$001818.4& 4.77 & ESI  &   7.9 & yes\\ 
SDSS~J023137.65$-$072854.4& 5.42 & ESI  &  26.2 & yes\\ 
SDSS~J033119.66$-$074143.1& 4.73 & ESI  &  17.8 & yes\\ 
SDSS~J075618.10$+$410409.0& 5.06 & ESI  &  10.8 & no \\ 
SDSS~J075907.57$+$180054.7& 4.82 & ESI  &  18.3 & yes\\ 
SDSS~J081333.30$+$350811.0& 4.92 & ESI  &  16.5 & no \\ 
SDSS~J082454.02$+$130217.0& 5.21 & ESI  &  22.5 & yes\\ 
SDSS~J083122.60$+$404623.0& 4.89 & ESI  &  19.6 & no \\ 
SDSS~J083429.40$+$214025.0& 4.50 & ESI  &  21.0 & no \\ 
SDSS~J083920.53$+$352459.3& 4.78 & ESI  &  13.9 & yes\\ 
SDSS~J095707.67$+$061059.5& 5.18 & ESI  &  18.1 & no \\ 
SDSS~J100449.58$+$404553.9& 4.87 & ESI  &  11.4 & no \\ 
SDSS~J100416.12$+$434739.0& 4.87 & ESI  &  19.7 & yes\\ 
SDSS~J101336.30$+$424027.0& 5.04 & ESI  &  22.7 & no \\ 
SDSS~J102833.46$+$074618.9& 5.15 & ESI  &  12.4 & no \\ 
SDSS~J104242.40$+$310713.0& 4.69 & ESI  &  24.8 & no \\ 
SDSS~J105445.43$+$163337.4& 5.15 & ESI  &  21.5 & yes\\ 
SDSS~J110045.23$+$112239.1& 4.73 & ESI  &  23.4 & yes\\ 
SDSS~J110134.36$+$053133.8& 5.04 & ESI  &  22.3 & yes\\ 
SDSS~J113246.50$+$120901.6& 5.18 & ESI  &  32.7 & yes\\ 
SDSS~J114657.79$+$403708.6& 5.00 & ESI  &  25.7 & yes\\ 
SDSS~J120036.72$+$461850.2& 4.74 & ESI  &  19.0 & yes\\ 
SDSS~J120110.31$+$211758.5& 4.58 & ESI  &  31.8 & yes\\ 
SDSS~J120207.78$+$323538.8& 5.30 & ESI  &  25.6 & yes\\ 
SDSS~J120441.73$-$002149.6& 5.09 & ESI  &  15.6 & yes\\ 
SDSS~J122042.00$+$444218.0& 4.66 & ESI  &  11.3 & no \\ 
SDSS~J122146.42$+$444528.0& 5.20 & ESI  &  15.5 & yes\\ 
SDSS~J123333.47$+$062234.2& 5.30 & ESI  &  14.1 & yes\\ 
SDSS~J124515.46$+$382247.5& 4.96 & ESI  &  16.4 & yes\\ 
SDSS~J125353.35$+$104603.1& 4.92 & ESI  &  23.8 & yes\\ 
SDSS~J130215.71$+$550553.5& 4.46 & ESI  &  24.0 & yes\\ 
SDSS~J131234.08$+$230716.3& 4.96 & ESI  &  19.2 & yes\\ 
SDSS~J133412.56$+$122020.7& 5.13 & ESI  &  10.8 & yes\\ 
SDSS~J134040.24$+$281328.1& 5.35 & ESI  &  23.0 & yes\\ 
SDSS~J134015.03$+$392630.7& 5.05 & ESI  &  17.7 & yes\\ 
SDSS~J141209.96$+$062406.9& 4.41 & ESI  &  25.6 & yes\\ 
SDSS~J141839.99$+$314244.0& 4.85 & ESI  &  15.3 & no \\ 
SDSS~J142103.83$+$343332.0& 4.96 & ESI  &  24.1 & no \\ 
SDSS~J143751.82$+$232313.3& 5.32 & ESI  &  24.4 & yes\\ 
SDSS~J143835.95$+$431459.2& 4.69 & ESI  &  27.0 & yes\\ 
SDSS~J144352.94$+$060533.1& 4.89 & ESI  &   5.6 & no \\ 
SDSS~J144331.17$+$272436.7& 4.42 & ESI  &  24.4 & yes\\ 
SDSS~J151320.89$+$105807.3& 4.62 & ESI  &   8.3 & yes\\ 
SDSS~J152345.69$+$334759.3& 5.33 & ESI  &   8.6 & no \\ 
SDSS~J153459.75$+$132701.4& 5.04 & ESI  &   4.9 & yes\\ 
SDSS~J153627.09$+$143717.1& 4.88 & ESI  &   8.2 & no \\ 
SDSS~J160734.22$+$160417.4& 4.79 & ESI  &  17.9 & yes\\ 
SDSS~J161425.13$+$464028.9& 5.31 & ESI  &  13.1 & yes\\ 
SDSS~J162626.50$+$275132.4& 5.26 & ESI  &  33.9 & yes\\ 
SDSS~J162629.19$+$285857.5& 5.04 & ESI  &  12.0 & yes\\ 
SDSS~J165436.80$+$222733.0& 4.68 & ESI  &  33.6 & no \\ 
SDSS~J165902.12$+$270935.1& 5.32 & ESI  &  23.1 & yes\\ 
SDSS~J173744.87$+$582829.6& 4.91 & ESI  &  12.9 & yes\\ 
SDSS~J221644.00$+$001348.0& 5.01 & ESI  &   8.2 & no \\ 
SDSS~J225246.43$+$142525.8& 4.88 & ESI  &  14.8 & yes\\ 
SDSS~J231216.40$+$010051.4& 5.07 & ESI  &   4.8 & no \\ 
\hline
\end{tabular}
\caption{\label{t_highres} Higher resolution spectra used in our
  analysis. Columns list the QSO name, R.A. and Dec. (J2000), emission
  redshift, the instrument used to take the spectrum, the median S/N
  per pixel over rest-frame wavelengths 1240--1280~\AA, and whether
  the QSO is in the GGG sample.}
\end{center}
\end{table}

\begin{table*}
\renewcommand{\arraystretch}{1}
\addtolength{\tabcolsep}{-2pt}
\scriptsize
\begin{center}
\begin{tabular}{ccccccccccc}
\hline
$z_{\rm JXP}$ & $\log_{10}N_{\rm JXP}$ & $z_{\rm NC}$ & $\log_{10}N_{\rm NC}$ &$z_{\rm hires}$ & $\log_{10}N_{\rm hires}$ & $z_{\rm best}$ & $\log_{10}N_{\rm best}$ & Label & Hi-res exists? &GGG?\\
\hline
 4.740 &  20.25 &  4.739 &  20.40 &  4.7395 &  20.30 &  4.7395 &  20.30 & J0040-0915 &y & y \\ 
 4.187 &  20.50 &    -   &    -   &    -    &    -   &  4.1888 &  20.60 & J0125-1043 &n & y \\ 
 4.887 &  20.75 &  4.886 &  20.70 &  4.8836 &  20.50 &  4.8836 &  20.50 & J0231-0728 &y & y \\ 
 4.658 &  20.95 &  4.657 &  21.00 &  4.6576 &  20.75 &  4.6576 &  20.75 & J0759+1800 &y & y \\ 
 4.098 &  21.05 &  4.096 &  21.05 &   -     &    -   &  4.0985 &  21.05 & J0800+3051 &n & y \\ 
  -    &    -   &  4.472 &  20.40 &  4.4720 &  20.30 &  4.4720 &  20.30 & J0824+1302 &y & y \\ 
 4.830 &  20.85 &  4.830 &  20.90 &  4.8305 &  20.75 &  4.8305 &  20.75 & J0824+1302 &y & y \\
 4.341 &  20.85 &  4.343 &  20.90 &  4.3441 &  20.60 &  4.3441 &  20.60 & J0831+4046 &y & n \\ 
 3.713 &  20.75 &  3.712 &  20.95 &  3.7100 &  20.75 &  3.7100 &  20.75 & J0834+2140 &y & n \\ 
 4.391 &  21.20 &  4.391 &  21.30 &  4.3920 &  21.15 &  4.3920 &  21.15 & J0834+2140 &y & n \\ 
 4.424 &  21.05 &  4.425 &  21.02 &   -     &    -   &  4.4227 &  21.05 & J0854+2056 &n & y \\ 
 4.795 &  20.45 &  4.794 &  20.45 &   -     &    -   &  4.7945 &  20.45 & J0913+5919 &n & y \\ 
 3.979 &  20.35 &   -    &    -   &   -     &    -   &  3.9790 &  20.35 & J0941+5947 &n & y \\ 
  -    &    -   &  4.862 &  20.40 &   -     &    -   &   -     &    -   & J0957+0519 &y & n \\ 
 4.473 &  20.40 &  4.472 &  20.55 &   -     &    -   &   -     &    -   & J1004+4347 &y & y \\ 
  -    &    -   &   -    &    -   &  4.4596 &  20.75 &  4.4596 &  20.75 & J1004+4347 &y & y \\ 
 4.798 &  20.55 &  4.805 &  20.50 &  4.7979 &  20.60 &  4.7979 &  20.60 & J1013+4240 &y & n \\ 
 4.257 &  20.70 &  4.259 &  20.30 &   -     &    -   &  4.2580 &  20.50 & J1023+6335 &n & y \\ 
 4.087 &  20.70 &  4.086 &  20.90 &  4.0861 &  20.75 &  4.0861 &  20.75 & J1042+3107 &y & n \\
  -    &    -   &   -    &    -   &  4.8165 &  20.70 &  4.8165 &  20.70 & J1054+1633 &y & y \\ 
  -    &    -   &   -    &    -   &  4.8233 &  20.50 &  4.8233 &  20.50 & J1054+1633 &y & y \\ 
 4.429 &  20.85 &   -    &    -   &   -     &    -   &   -     &    -   & J1100+1122 &y & y \\
 4.397 &  21.60 &  4.395 &  21.55 &  4.3954 &  21.65 &  4.3954 &  21.65 & J1100+1122 &y & y \\ 
 4.346 &  21.40 &  4.347 &  21.35 &  4.3441 &  21.35 &  4.3441 &  21.35 & J1101+0531 &y & y \\ 
 4.380 &  21.20 &   -    &    -   &  4.3801 &  21.15 &  4.3801 &  21.15 & J1132+1209 &y & y \\ 
 5.015 &  20.75 &  5.015 &  20.60 &  5.0165 &  20.70 &  5.0165 &  20.70 & J1132+1209 &y & y \\ 
 4.476 &  20.60 &  4.476 &  20.65 &  4.4767 &  20.45 &  4.4767 &  20.45 & J1200+4618 &y & y \\ 
 3.799 &  21.35 &  3.807 &  21.20 &  3.7961 &  21.25 &  3.7961 &  21.25 & J1201+2117 &y & y \\ 
 4.156 &  20.60 &   -    &    -   &  4.1579 &  20.50 &  4.1579 &  20.50 & J1201+2117 &y & y \\ 
 4.793 &  20.75 &  4.798 &  20.75 &  4.7956 &  21.10 &  4.7956 &  21.10 & J1202+3235 &y & y \\ 
 4.811 &  20.75 &   -    &    -   &  4.8106 &  20.75 &  4.8106 &  20.75 & J1221+4445 &y & y \\ 
 4.926 &  20.35 &  4.931 &  20.70 &  4.9311 &  20.55 &  4.9311 &  20.55 & J1221+4445 &y & y \\ 
 4.711 &  20.50 &   -    &    -   &   -     &    -   &   -     &    -   & J1233+0622 &y & y \\ 
 4.448 &  20.80 &  4.447 &  20.70 &  4.4467 &  20.45 &  4.4467 &  20.45 & J1245+3822 &y & y \\ 
 4.213 &  20.50 &  4.213 &  20.40 &   -     &    -   &  4.2130 &  20.45 & J1301+2210 &n & y \\ 
 3.937 &  21.10 &  3.937 &  21.10 &   -     &    -   &  3.9387 &  21.10 & J1309+1657 &n & y \\ 
 4.303 &  20.55 &  4.303 &  20.50 &   -     &    -   &  4.3027 &  20.52 & J1332+4651 &n & y \\ 
   -   &    -   &   -    &    -   &  4.7636 &  20.35 &  4.7636 &  20.35 & J1334+1220 &y & y \\ 
 4.348 &  20.55 &  4.348 &  20.50 &   -     &    -   &  4.3480 &  20.52 & J1337+4155 &n & y \\ 
 5.003 &  20.85 &   -    &    -   &   -     &    -   &    -    &    -   & J1340+2813 &y & y \\ 
   -   &    -   &  5.096 &  20.30 &   -     &    -   &    -    &    -   & J1340+2813 &y & y \\ 
 4.826 &  21.05 &  4.827 &  21.05 &  4.8258 &  21.20 &  4.8258 &  21.20 & J1340+3926 &y & y \\ 
 4.109 &  20.35 &   -    &    -   &  4.1093 &  20.35 &  4.1093 &  20.35 & J1412+0624 &y & y \\ 
  -    &    -   &  4.322 &  20.40 &   -     &    -   &    -    &    -   & J1418+3142 &y & n \\ 
 3.958 &  20.55 &   -    &    -   &   -     &    -   &  3.9628 &  21.00 & J1418+3142 &y & n \\ 
 4.453 &  20.35 &  4.453 &  20.45 &   -     &    -   &    -    &    -   & J1418+3142 &y & n \\ 
 4.114 &  20.60 &  4.112 &  20.70 &   -     &    -   &  4.1140 &  20.65 & J1420+6155 &n & y \\ 
   -   &    -   &  4.665 &  20.35 &  4.6644 &  20.30 &  4.6644 &  20.30 & J1421+3433 &y & n \\ 
 4.093 &  20.30 &   -    &    -   &   -     &    -   &  4.0929 &  20.30 & J1427+3308 &n & y \\ 
 4.526 &  20.60 &  4.527 &  20.60 &   -     &    -   &  4.5218 &  20.60 & J1436+2132 &n & y \\ 
 4.800 &  21.10 &  4.801 &  21.10 &  4.8007 &  21.20 &  4.8007 &  21.20 & J1437+2323 &y & y \\ 
 4.400 &  20.80 &  4.398 &  20.85 &  4.3989 &  20.80 &  4.3989 &  20.80 & J1438+4314 &y & y \\ 
  -    &    -   &  4.355 &  20.35 &   -     &    -   &   -     &    -   & J1443+0605 &y & n \\ 
 4.223 &  20.95 &  4.222 &  21.05 &  4.2237 &  20.95 &  4.2237 &  20.95 & J1443+2724 &y & y \\ 
 4.088 &  21.45 &  4.089 &  21.57 &   -     &    -   &  4.0885 &  21.51 & J1511+0408 &n & y \\ 
 4.304 &  21.05 &  4.305 &  21.10 &   -     &    -   &  4.3043 &  21.08 & J1524+1344 &n & y  \\ 
 3.818 &  20.45 &  3.817 &  20.30 &   -     &    -   &  3.8175 &  20.38 & J1532+2237 &n & y \\ 
   -   &    -   &  4.466 &  20.30 &  4.4740 &  20.40 &  4.4740 &  20.40 & J1607+1604 &y & y \\ 
 4.915 &  20.90 &  4.912 &  21.00 &  4.9091 &  21.00 &  4.9091 &  21.00 & J1614+4640 &y & y \\ 
 4.462 &  20.70 &  4.462 &  20.85 &   -     &    -   &   -     &    -   & J1626+2751 &y & y \\
 4.312 &  21.20 &  4.313 &  21.30 &  4.3105 &  21.30 &  4.3105 &  21.30 & J1626+2751 &y & y \\ 
 4.498 &  20.95 &  4.498 &  21.00 &  4.4973 &  21.05 &  4.4973 &  21.05 & J1626+2751 &y & y \\ 
 4.605 &  20.55 &   -    &    -   &  4.6067 &  20.55 &  4.6067 &  20.55 & J1626+2858 &y & y \\ 
 4.083 &  20.60 &  4.082 &  20.50 &   -     &    -   &  4.0825 &  20.55 & J1634+2153 &n & y \\ 
  -    &    -   &  4.101 &  20.60 &   -     &    -   &   -     &    -   & J1654+2227 &y & n \\ 
 4.001 &  20.60 &  4.003 &  20.75 &  4.0023 &  20.55 &  4.0023 &  20.55 & J1654+2227 &y & n \\ 
 4.742 &  20.70 &  4.740 &  20.60 &  4.7424 &  20.80 &  4.7424 &  20.80 & J1737+5828 &y & y \\ 
  -    &    -   &   -    &    -   &  4.7475 &  20.55 &  4.7475 &  20.55 & J2252+1425 &y & y \\ 
 4.257 &  21.10 &  4.256 &  20.80 &   -     &    -   &   -     &    -   & J2312+0100 &y & n \\ 
\hline
\end{tabular}
\caption{\label{t_dla} DLAs identified in the GGG spectra and and
  other higher-resolution spectra. The first four columns list the
  redshift and \NHI\ estimate in the GMOS-resolution spectra by two of
  us (JXP and NHMC).  The fifth and sixth columns give measurements
  from a high-resolution spectrum of the QSO, if one exists. The
  seventh and eight columns give the `best' estimate of for the DLA
  redshift and \NHI. This is value from the high-resolution spectrum
  if one exists, otherwise it is the mean of the estimates from JXP
  and NHMC. In the first eight columns, a dash means no DLA was
  identified. The ninth column gives the QSO name, and the tenth
  column lists whether the QSO has high-resolution spectrum. The
  last column lists whether the QSO is part of the GGG sample used
  to measure \OmHI.}
\end{center}
\end{table*}

\begin{table}
\renewcommand{\arraystretch}{1}
\addtolength{\tabcolsep}{-2pt}
\footnotesize
\begin{center}
\begin{tabular}{lccc}
\hline
Name & $z_\mathrm{min}$ & $z_\mathrm{max}$ & $z_\mathrm{qso}$ \\
\hline
SDSS~J001115.23+144601.8 & 4.037 & 4.870 & 4.970 \\
SDSS~J004054.65-091526.8 & 4.046 & 4.880 & 4.980 \\
SDSS~J010619.24+004823.3 & 3.598 & 4.358 & 4.449 \\
SDSS~J012509.42-104300.8 & 3.639 & 4.406 & 4.498 \\
SDSS~J021043.16-001818.4 & 3.868 & 4.674 & 4.770 \\
SDSS~J023137.65-072854.4 & 4.417 & 5.313 & 5.420 \\
SDSS~J033119.66-074143.1 & 3.838 & 4.638 & 4.734 \\
SDSS~J033829.30+002156.2 & 4.096 & 4.939 & 5.040 \\
SDSS~J073103.12+445949.4 & 4.061 & 4.898 & 4.998 \\
SDSS~J075907.57+180054.7 & 3.911 & 4.723 & 4.820 \\
SDSS~J080023.01+305101.1 & 3.789 & 4.581 & 4.676 \\
SDSS~J080715.11+132805.1 & 3.961 & 4.782 & 4.880 \\
SDSS~J081806.87+071920.2 & 3.746 & 4.531 & 4.625 \\
SDSS~J082212.34+160436.9 & 3.649 & 4.418 & 4.510 \\
SDSS~J082454.02+130217.0 & 4.237 & 5.103 & 5.207 \\
\end{tabular}
\caption{\label{t_zpath} The start and end redshifts for each QSO used
  to calculate the redshift search path for DLAs. This is a stub; the
  full table is available online.}
\end{center}
\end{table}

\clearpage

\appendix
\section{Mock spectra}
\label{a_mocks}

We generated a set of mock spectra to quantify the reliability and
completeness of our DLA candidates in the low-resolution GMOS
spectra. Here we describe how these mocks were produced.

One mock spectrum was generated for each real GMOS spectrum, assuming
the same noise properties and the same QSO redshift. Therefore the
sample of mocks has the same redshift and S/N distribution to the real
GGG spectra. We model the forest absorption by a distribution of Voigt
profiles. Due the difficulty of profile-fitting the strongly absorbed
\lya\ forest at high redshifts, the $\NHI$, $b$ and $z$ distribution
of \lya\ forest lines at $z > 4$ is not well known. However the
distribution at $z\sim2.5$ has been measured \citep[e.g.][]{Kim13,
  Rudie13}. Therefore we assume the shape of $f(\NHI)$ at
$z\sim4$--$5$ is the same as that used by \citet{Prochaska14_IGM} at
$z\sim2.5$, and increase its normalisation until the mean flux of the
mock spectra at $z=4.5$ matches the value from
\citet{Becker13_taueff}. DLAs were generated using $f(\NHI,X)$ from
\citet{OMeara13}, and we assume $f(\NHI)$ is
redshift-independent, whereas $f(\NHI,X)$ evolves as $(1+z)^{1.5}$.

We initially did not include any line clustering in the \lya\ forest,
but found that this produced spectra which were markedly different to
the real spectra: there were too few regions with very strong
absorption and also too few regions with low absorption. To address
this we introduced line clustering, similar to that used by
\citet{Liske08} to model the \lya\ forest at $z\sim3$. This involves
generating absorption at `clump' positions rather than individual
lines. For each clump, 0, 1 or more lines are produced, with the
number taken from a Borel distribution \citep{Saslaw89} with
$\beta=0.6$. Each line in a clump is offset from the clump redshift by
a velocity drawn from a Gaussian distribution with
$\sigma=250$~\kms. These values of $\beta$ and $\sigma$ were chosen by
a parameter grid search, varying each until values were found which
produce in mock spectra with a \lya\ forest which match the flux
distribution of the real GMOS spectra. The number of clumps was set
such that the mean transmission in the \lya\ forest matches the
effective optical depth at $z=4.5$ derived by \citet{Becker13_taueff}.

We then generated a QSO continuum from the PCA components presented by
\citet{Suzuki05}, derived using a sample of low redshift QSOs observed
with the UV Faint Object Spectrograph. We set the QSO redshift to that
of the matching GMOS QSO, and added noise to the mock using the same
noise array as the GMOS spectrum, normalised so that the median S/N of
the mock and the real spectra in the range
$7600$--$7800$~\AA\ matches.  Using the noise array from the real
spectra for the mocks is an approximation, as the noise properties
vary with the QSO spectrum (strong absorbers and strong emission lines
affect the noise level). However, the variations in noise due to these
effects is small in the \lya\ forest, so we believe this is a good
approximation.

\begin{figure}
\includegraphics[width=1.0\columnwidth]{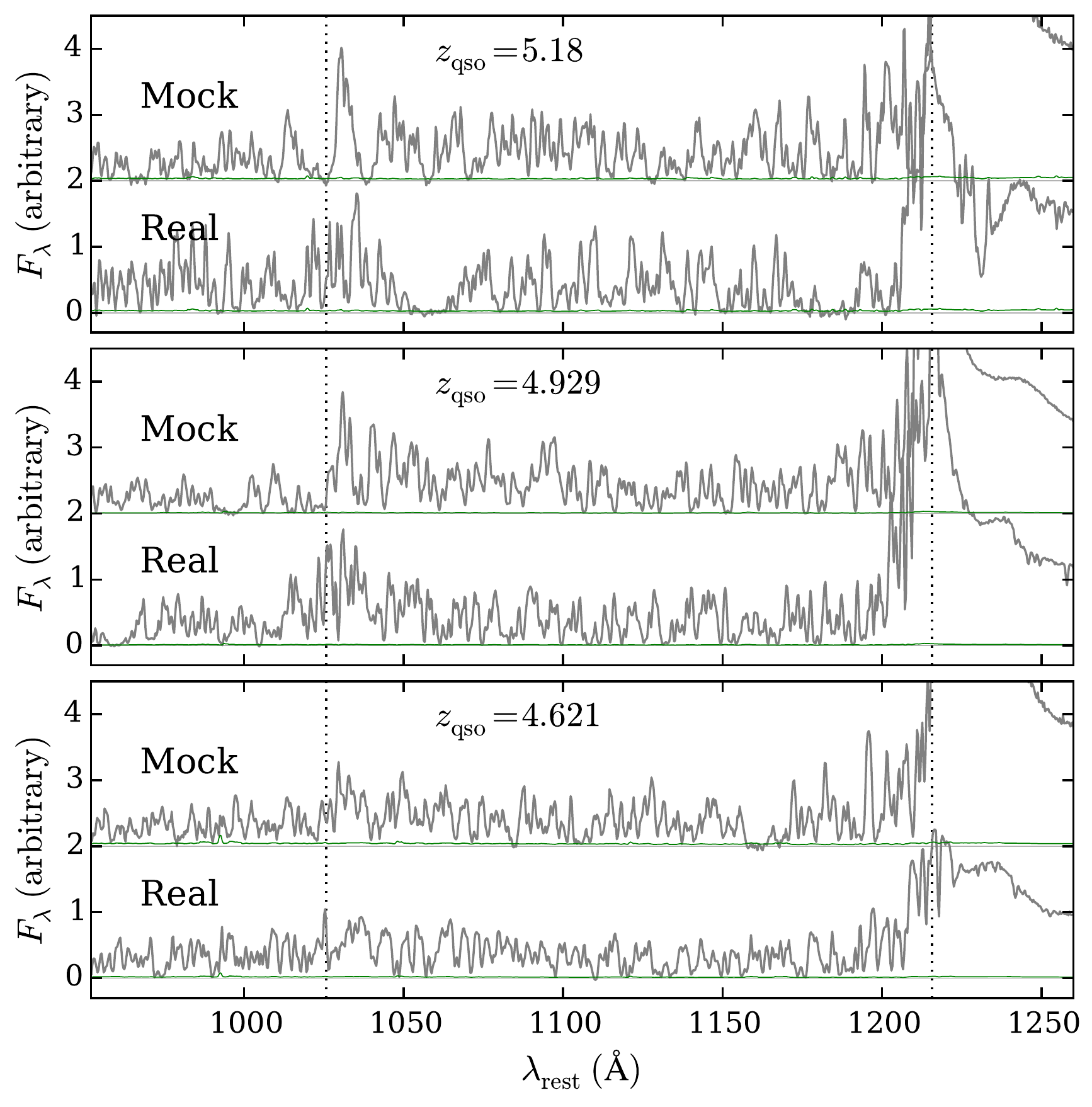}
\caption{\label{f_cf_mock_real} Three mock GMOS spectra, selected at
  random, with their corresponding real spectra. The real and mock
  spectra are normalised in the rest frame wavelength region
  $940$--$1200$~\AA\ and offset for clarity. The flux distribution in
  the \lya\ forest (between the two dotted vertical lines), where we
  search for DLAs, is very similar. The thin green lines show the
  $1\sigma$ error array.}
\end{figure}

Figure~\ref{f_cf_mock_real} shows three example mock spectra and their
corresponding real spectra, selected at random from our sample. The
\lya\ forest distribution in the mocks matches closely the
distribution seen in the real spectra. We do not expect these mocks to
correctly reproduce the mean optical depth at the Lyman limit or the
power spectrum of \lya\ flux absorption. However, our aim is not to
reproduce all properties of the real spectra. Instead we aim to create
mock spectra which match by eye the \lya\ forest at GMOS resolution,
the most important characteristic for DLA identification.

We did not include metal absorption in the mocks. The similarity
between the mocks and the real spectra, and the agreement between the
correction factors $k_\mathrm{real}$ and $k_\mathrm{found}$ derived
from the mocks and high-resolution spectra suggest their inclusion is
unnecessary.

\subsection{High \NHI\ DLAs}

\begin{figure}
\begin{center}
\includegraphics[width=0.95\columnwidth]{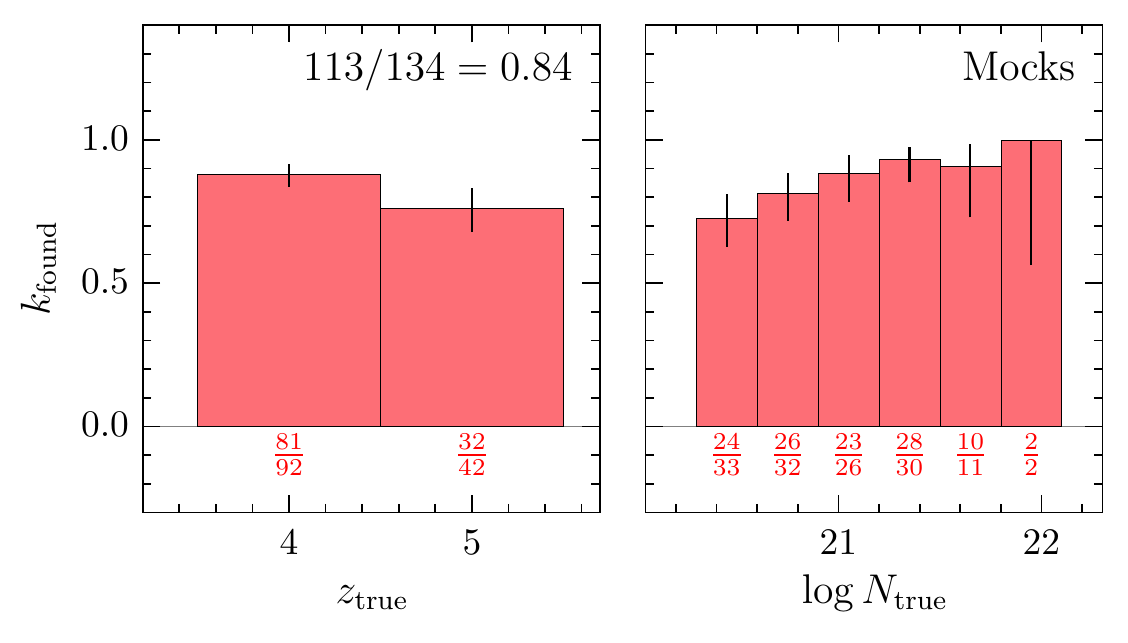}
\caption{\label{f_kfound_hiN} The fraction of true DLAs that were
  correctly identified by one of the authors (NHMC),
  $k_\mathrm{found}$, as a function of the true redshift and
  \NHI\ found using the mock spectra. This includes the mock
  sightlines with additional strong \NHI\ DLAs.}
\end{center}
\end{figure}

\begin{figure}
\begin{center}
\includegraphics[width=0.95\columnwidth]{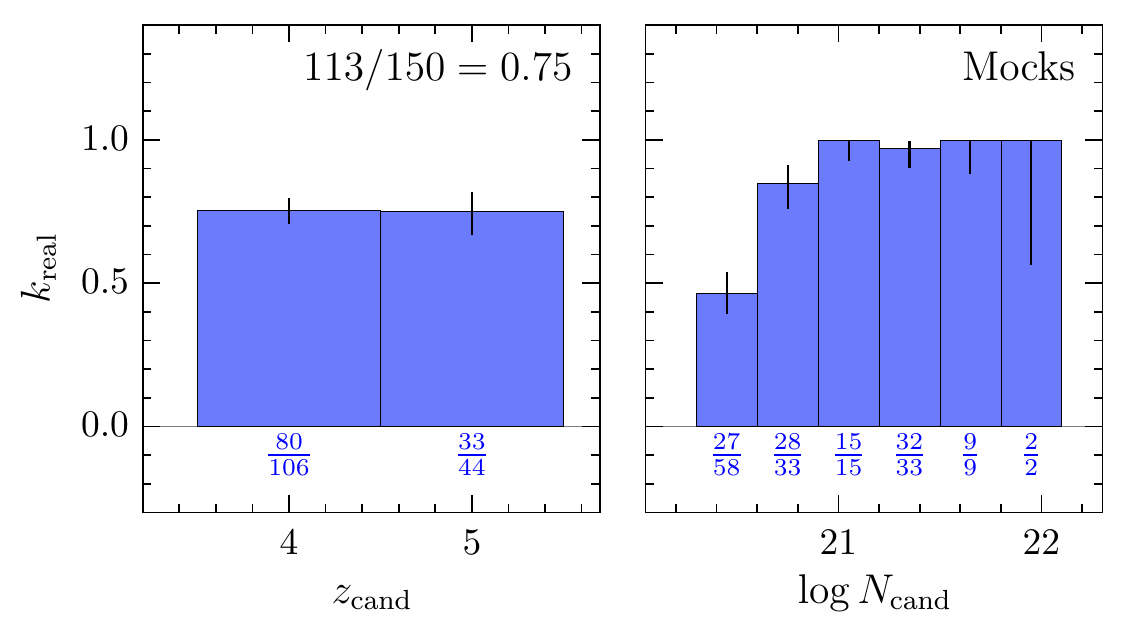} 
\caption{\label{f_kreal_hiN} The fraction of non-spurious DLA
  candidates, $k_\mathrm{real}$ by one of the authors (NHMC), as a
  function of the candidate redshift and \NHI\ for the mock
  spectra. This includes the mock sightlines with additional strong
  \NHI\ DLAs.}
\end{center}
\end{figure}

DLAs in the column density range $\NHI=10^{21-21.8}$~\cmm\ make the
dominant contribution to \OmHI, and it is thus important to correctly
measure the uncertainty in $k_\mathrm{real}$ and $k_\mathrm{found}$
for this \NHI\ range. There are only $\sim 10$ DLAs in this column
density range in both the mocks and the high-resolution sample, so the
uncertainties in this correction are large. Therefore we generated
further mocks with an enhanced incidence rate of high
\NHI\ systems. We did this by generating 10 times more mocks than were
used above, using the same line distribution. Due to time constraints,
we were unable to search by eye every one of these mocks. Instead we
selected just 100 spectra: the 50 containing the highest \NHI\ DLAs,
and a further 50 selected at random from the remainder. This formed a
sample of 100 new mock spectra which we searched for high
\NHI\ systems. 50 were included without requiring a DLA to present so
that when scanning the spectra by eye, the searcher would not be
certain that every spectrum contains a DLA. The $k_\mathrm{found}$,
$k_\mathrm{real}$ values found by including these extra sightlines
into our mock sample are shown in Figures \ref{f_kfound_hiN} and
\ref{f_kreal_hiN}. These show that the probability of a spurious DLA
at $\NHI>10^{21}$~\cmm\ is just 1\%--5\%, using binomial statistics
with

\begin{figure}
\begin{center}
\includegraphics[width=1.01\columnwidth]{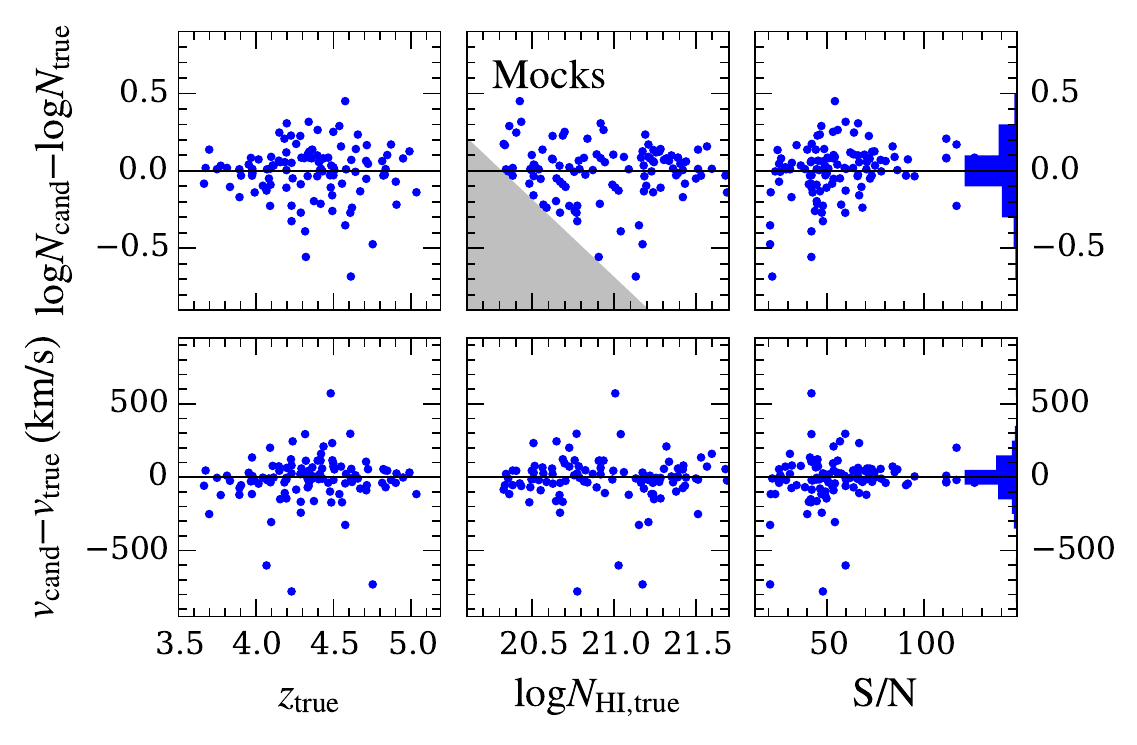}
\caption{\label{f_match_hiN} The difference between \NHI\ estimated
  for DLA candidates in low resolution spectra ($N_\mathrm{HI,cand}$)
  and $N_\mathrm{HI,true}$ measured from mock linelists by one of the
  authors (NHMC), including the extra sightlines with additional
  strong \NHI\ DLAs. This shows there is no strong systematic offset
  in the estimated \NHI\ as a function of redshift, even for
  $\NHI\sim10^{21.5}$~\cmm\ DLAs.}
\end{center}
\end{figure}

The $\log \NHI$ and velocity differences between the candidate and
true values are shown in Figure \ref{f_match_hiN}. This shows that
even at high \NHI, there is no strong systematic offset from the true
value.


\end{document}